\documentclass[manuscript,screen]{acmart}
\usepackage{pifont}
\usepackage{multirow}
\usepackage{array}
\usepackage{booktabs}
\usepackage{tabularx}
\usepackage{threeparttable}
\usepackage{siunitx}
\usepackage{footmisc}
\AtBeginDocument{%
  }

\setcopyright{acmlicensed}
\copyrightyear{2018}
\acmYear{2018}
\acmDOI{XXXXXXX.XXXXXXX}
\acmConference[Conference acronym 'XX]{Make sure to enter the correct
  conference title from your rights confirmation email}{June 03--05,
  2018}{Woodstock, NY}
\acmISBN{978-1-4503-XXXX-X/2018/06}

\begin{document}

\title{Membership Inference Attacks on Recommender Systems: A Survey}

\author{Jiajie He}
\authornotemark[1]
\affiliation{%
  \institution{University of Maryland, Baltimore County}
  \city{Baltimore}
  \country{USA}}
\email{jiajieh1@umbc.edu}

\author{Xintong Chen}
\authornotemark[1]
\affiliation{%
  \institution{McMaster University}
  \city{Hamilton}
  \country{Canada}}
\email{chenx545@mcmaster.ca}
\footnote{Both authors contributed equally to this research.}

\author{Wenjin Liu}
\affiliation{%
  \institution{Nanyang Technological University}
  \city{Singapore}
  \country{SG}}
  
\author{Xinyang Fang}
\affiliation{%
  \institution{University of Southern California}
  \city{Los Angeles}
  \country{USA}}
  
\author{Min-chun Chen}
\affiliation{%
  \institution{University of Maryland, Baltimore County}
  \city{Baltimore}
  \country{USA}}

\author{Yuechun Gu}
\affiliation{%
  \institution{University of Maryland, Baltimore County}
  \city{Baltimore}
  \country{USA}}
  
\author{Keke Chen}
\affiliation{%
  \institution{University of Maryland, Baltimore County}
  \city{Baltimore}
  \country{USA}}
\renewcommand{\shortauthors}{He et al.}

\begin{abstract}
Recommender systems (RecSys) have been widely applied across E-commerce, finance, healthcare, and social media and have become increasingly influential in shaping user behavior and decision-making, underscoring their growing impact across domains. Since RecSys heavily relies on user data, its privacy concerns are significant and need to be addressed urgently.
Recent studies on membership inference attacks (MIAs) in RecSys highlight this need. MIAs aim to infer whether a user or an interaction record was used to train a target ReSys model. The success of MIA can lead to severe privacy breaches, e.g., inferring users' special lifestyles. MIAs in RecSys have features distinct from other MIAs in classification models or large language models. However, no systematic survey on this topic has yet been conducted. We present the first comprehensive survey on RecSys MIAs, exploring their taxonomy, design principles, evaluation methods, and defense mechanisms. Based on the summary of existing studies in this area, we also outline several promising future research directions. This survey will raise awareness of privacy risks among RecSys researchers, practitioners, and users, and promote privacy protection practices in RecSys design.  
\end{abstract}



\keywords{Do, Not, Use, This, Code, Put, the, Correct, Terms, for,
  Your, Paper}

\received{20 February 2007}
\received[revised]{12 March 2009}
\received[accepted]{5 June 2009}

\maketitle

\section{Introduction}
\label{sec:Introduction}

Recommender systems (RecSys) have experienced significant advances over the past decade and are widely deployed across various applications, including job matching~\cite{10.1145/3543507.3583355}, e-commerce~\cite{chen2023knowledgegraphcompletionmodels}, entertainment~\cite{gao2023chatrecinteractiveexplainablellmsaugmented}, and social media~\cite{he2020lightgcn}. In addition to advances in recommendation algorithms and computational resources, the availability of large-scale user data has been a key factor driving their success~\cite{Zhao_2024}. RecSys learn user preferences from diverse behavioral information, including purchase histories, browsing activities, ratings, search queries, and social interactions. Unlike conventional classification models, RecSys typically learn from sparse and heterogeneous user--item interactions, where individual records are connected through shared users, items, and collaborative representations. While these characteristics enable highly personalized recommendations\cite{he2017neural,he2020lightgcn,8594844,10.1145/3404835.3463032}, they also raise significant privacy concerns\cite{10.1145/3705328.3748052,11372209,zhu2023membership,10.1145/3773078.3831733,10.1145/3568954}, as sensitive user information may be encoded in learned representations and reflected in recommendation outputs.

Among various privacy threats\cite{zhang2020,shokri17,melis2019exploiting,carlini2022lira,NEURIPS2019_60a6c400} in machine learning, \textbf{membership inference attacks (MIAs)} have received substantial attention due to their ability to reveal information about the data used during model training. Given a trained target model and a candidate data record, an MIA aims to determine whether the record was included in the model's training dataset. Shokri et al.~\cite{shokri17} conducted one of the earliest systematic studies of MIAs, demonstrating that an adversary could infer training-data membership using prediction outputs from a target classification model, even without direct access to its parameters or training data. Subsequent studies have developed more sophisticated attack strategies by exploiting prediction confidence and entropy\cite{salem2018mlleaksmodeldataindependent,song2021systematic}, training loss and loss trajectories\cite{Yeom18,sablayrolles2019white,liu2022membershipinferenceattacksexploiting}, likelihood ratios\cite{carlini2022lira,pmlr-v235-zarifzadeh24a}, and other membership-dependent signals  such as gradient information~\cite{Nasr_2019} and
decision-boundary characteristics ~\cite{Choquette-Choo21}. MIAs have since been investigated across diverse machine learning domains, including computer vision~\cite{Nasr_2019,Choquette-Choo21,carlini2022lira}, natural language processing~\cite{wen2024membershipinferenceattacksincontext,mattern-etal-2023-membership,shi2024detectingpretrainingdatalarge}, and audio processing~\cite{shah21_interspeech,TEIXEIRA2026101812,tseng22_interspeech}. These studies demonstrate that learned models may reveal information about their training data through differences in their behavior on member and non-member samples. Such leakage can become particularly sensitive when membership in a training dataset is associated with private personal information.

Recent studies have extended membership inference to recommender systems, demonstrating that RecSys models are also vulnerable to membership privacy leakage~\cite{10.1145/3705328.3748052, 10.1145/3773078.3831733,zhu2023membership,chi2024shadowfreemembershipinferenceattacks,yuan2023interaction,zhong2024interaction}. However, directly applying conventional MIA techniques to RecSys introduces additional challenges arising from their structured interaction data, collaborative learning mechanisms, and personalized recommendation outputs. Unlike conventional classification models, which often produce prediction labels or class probabilities, RecSys typically generate ranked lists of items based
on user preferences, item characteristics, and historical interactions. In practical black-box settings, attackers may only observe top-$K$ recommendation lists, without access to prediction scores, model parameters, or training losses~\cite{zhang2021membership}. Consequently, membership inference may require indirect signals derived from recommendation rankings, the relationships between historical and recommended items, or item embeddings estimated using auxiliary data.
However, the internal item embeddings learned by the target recommender are often inaccessible to black-box attackers, requiring them to construct approximate
representations that may differ from those learned by the target model. Such representation estimation errors can introduce bias into similarity-based membership
signals and undermine the reliability of membership inference~\cite{wang2022debiasing}. Furthermore, these indirect signals do not necessarily indicate membership, as similar recommendation patterns may also arise from shared user preferences, item popularity, or collaborative generalization~\cite{chi2024shadowfreemembershipinferenceattacks}.

\begin{figure*}[t]
    \centering
    \includegraphics[width=\textwidth]{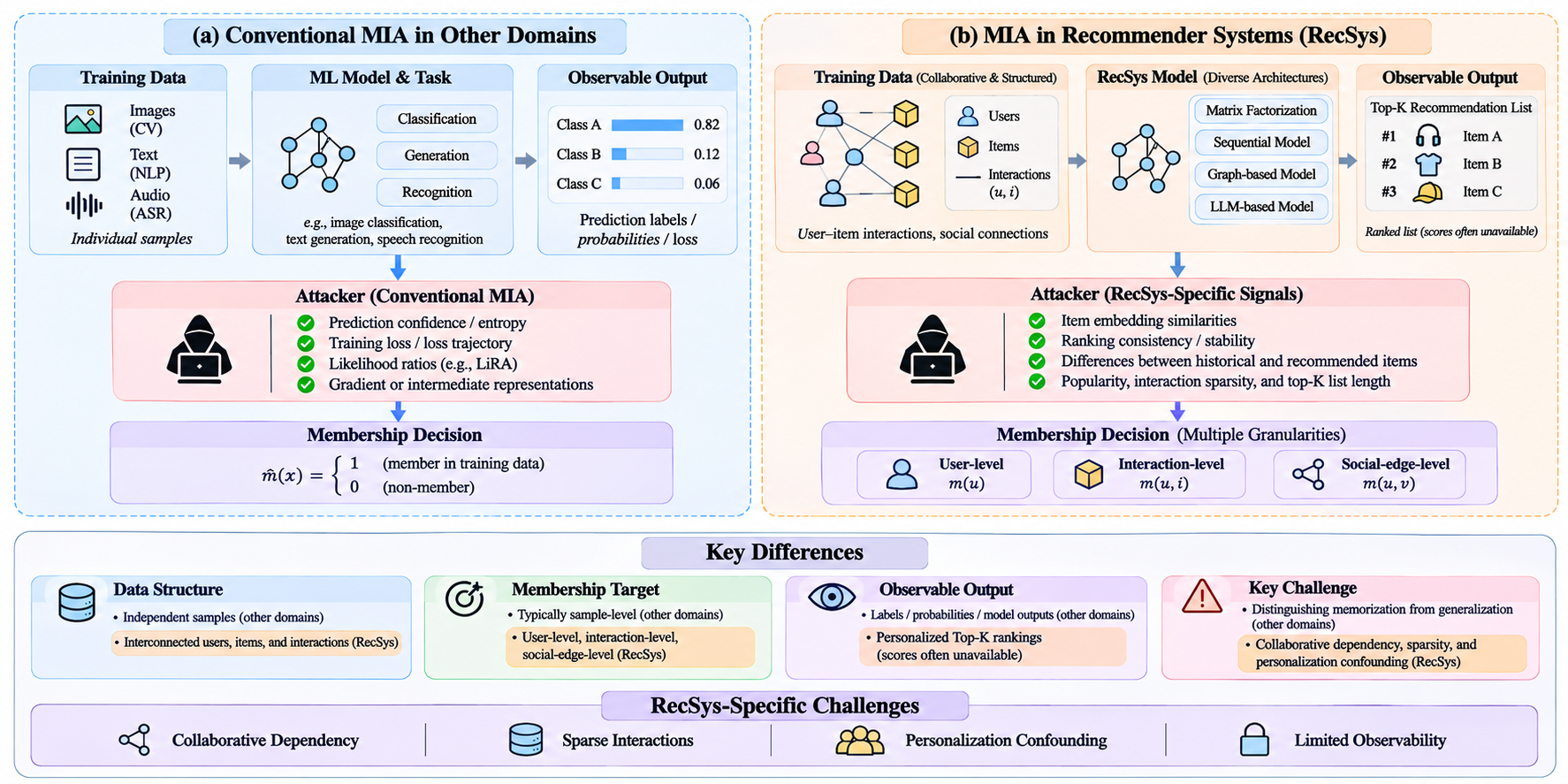}
    \caption{Comparison between conventional MIAs and MIAs in recommender systems, highlighting differences in training data, observable outputs, membership targets, attack signals, and inference challenges.}
    \label{fig:conventional-vs-recsys-mia}
\end{figure*}

As illustrated in Fig.~\ref{fig:conventional-vs-recsys-mia}, the distinction between conventional MIAs and RecSys MIAs extends beyond the type of observable output
to the structure of training data, the granularity of membership targets, and the mechanisms through which training information influences model behavior. These differences motivate the following RecSys-specific considerations.

\textbf{Unique Features of MIAs in RecSys.}
Several characteristics of recommender systems introduce distinct challenges and opportunities for membership inference:

\begin{itemize}

\item \textbf{Diverse Membership Targets.}
Unlike conventional MIAs that typically investigate the membership of individual training samples, RecSys contain privacy-sensitive information at multiple interconnected granularities. User-level MIAs determine whether a user's interaction history contributed to model training ~\cite{zhang2021membership,
zhu2023membership, chi2024shadowfreemembershipinferenceattacks}, whereas interaction-level MIAs examine whether a specific user--item interaction was included in the training data ~\cite{yuan2023interaction,zhong2024interaction,
10.1145/3705328.3748052}. In social RecSys, membership inference can further
target whether a particular user--user relationship was included in the training graph ~\cite{10.1145/3726302.3730086}.

These membership targets introduce different inference challenges because their contributions to model training occur at different levels. User-level membership may involve an entire interaction history, whereas interaction-level membership requires identifying the influence of a single interaction in the presence
of other interactions from the same user. Similarly, social relation-level membership requires isolating the contribution of a particular social edge from other behavioral and social relationships. Therefore, the choice of membership granularity affects the construction of attack samples, the interpretation
of membership signals, and the evaluation of inference accuracy. Importantly, inferring whether a relationship was included in the training graph should be
distinguished from inferring whether that relationship exists, as the latter constitutes a broader link inference problem.

\item \textbf{Recommendation-Specific Attack Signals.}
Conventional MIAs often exploit prediction confidence, training loss, or other model-dependent statistics. In contrast, RecSys frequently expose only personalized top-$K$ recommendation lists, making such information inaccessible in ranking-only black-box settings. Consequently, attackers must infer membership from indirect signals, such as the similarity between historical and recommended items, ranking patterns, and item popularity ~\cite{zhang2021membership, chi2024shadowfreemembershipinferenceattacks}

Extracting reliable membership signals is further complicated by several recommendation-specific factors.
Interaction sparsity limits the historical information available for estimating user preferences, while
long-tailed item distributions create uneven information availability across items. Moreover, top-$K$ recommendation interfaces expose only a restricted portion of the model's ranking, potentially concealing membership-dependent patterns. When embedding-based signals are employed, attackers
may additionally need to approximate inaccessible item embeddings, introducing estimation bias
~\cite{wang2022debiasing}. Together, these characteristics make it challenging to identify membership-dependent signals that remain reliable despite incomplete observations, representation uncertainty, and variations in user--item interactions.

\item \textbf{Collaborative Data Dependencies.}
Unlike learning settings where training samples can be treated as relatively independent observations,
many RecSys learn preferences from interconnected user--item interactions. Collaborative filtering,
shared item embeddings, and graph-based representations allow information from one user's interactions to
influence the representations and recommendations associated with other users.

This interdependence complicates membership inference because the recommendation output for a target user
may reflect both the user's own training interactions and preferences learned from other users.
For example, two users with similar historical interests may receive similar recommendations even
when only one user's interactions were included in model training. Likewise, highly popular items
may appear in recommendation lists regardless of whether the target user's data contributed to
training. Consequently, similarity between historical interactions and recommendation outputs cannot always be attributed to membership leakage \cite{chi2024shadowfreemembershipinferenceattacks}.

This creates a fundamental attribution challenge: an attacker must distinguish recommendation behavior
caused by the target's training participation from behavior arising through collaborative generalization,
item popularity, or personalization mechanisms. Moreover, differences between shadow and target
recommenders can further complicate the estimation of membership-dependent signals, as demonstrated
by debiasing-based membership inference research ~\cite{wang2022debiasing}. These dependencies suggest that reliable privacy evaluation should account for user similarity, interaction distributions, and recommendation policies when comparing members and non-members, rather than attributing all observed recommendation differences directly to training membership.

\item \textbf{Heterogeneous Recommendation Architectures.}
Modern RecSys employ diverse architectures, including matrix-factorization-based, sequential, graph-based, federated, cross-domain, and large language model (LLM)-based approaches. These architectures differ in how user information is represented and propagated, leading to distinct membership leakage mechanisms and attacker capabilities.

For instance, latent representations, sequential dependencies, and graph-based information propagation
introduce different sources of membership signals ~\cite{zhang2021membership,zhu2023membership}. Federated RecSys may expose membership information through shared model updates~\cite{yuan2023interaction}, while cross-domain RecSys introduce potential leakage through knowledge transfer across domains ~\cite{11036029}. LLM-based RecSys further introduce membership inference over inference-time contextual examples, which differs from conventional training-data membership ~\cite{10.1145/3773078.3831733}.

These architectural differences complicate the transferability of existing MIAs across recommendation settings and necessitate architecture-aware attack design and privacy evaluation.
\end{itemize}

\textbf{Direct Impacts on Individuals.}
Membership leakage in RecSys can have substantial privacy implications because user interactions often reflect sensitive personal behaviors and preferences. For example, determining whether a specific purchase record was included in an e-commerce recommender's training data may reveal information about a user's consumption habits. In healthcare recommendation, membership inference involving sensitive medical items could expose treatment-related information when combined with auxiliary knowledge. Similarly, inferring private social relationships incorporated into recommendation models may reveal otherwise hidden connections between individuals. Unlike general user-level membership, which may only indicate participation in a service, interaction-level membership can reveal more specific behavioral information. Therefore, the privacy consequences of RecSys MIAs depend on both the granularity of the inferred membership and the sensitivity of the underlying user information. These risks highlight the importance of investigating membership privacy at different levels rather than treating all recommendation data as equivalent.

\textbf{Motivation and Scope of This Survey.}
Existing surveys, such as Hu et al.~\cite{hu2022}, have systematically reviewed membership inference attacks and defense mechanisms in general machine learning settings. However, these surveys do not comprehensively address how recommendation-specific data structures, ranking mechanisms, collaborative dependencies, and heterogeneous architectures influence membership leakage. Recent research has expanded RecSys MIAs from user-level attacks to interaction-level and social-relation-level inference, while emerging paradigms such as federated, cross-domain, and LLM-based recommendation introduce additional attack surfaces. These developments motivate a dedicated survey that goes beyond transferring conventional MIA taxonomies to recommendation models. In particular, understanding what constitutes membership in different recommendation settings, which observable signals enable its inference, and how architectural characteristics affect attack effectiveness is essential for systematically evaluating privacy risks in RecSys.

To address these challenges, we provide a systematic review of membership inference attacks and privacy protection mechanisms specifically designed for recommender systems. We organize existing research from multiple perspectives, including membership targets, adversarial knowledge, recommendation architectures, and information exposure mechanisms. Furthermore, we examine the relationships between attack methodologies, evaluation protocols, and defense strategies to identify the limitations of existing approaches and opportunities for future research. The main contributions of this survey are summarized as follows:

\begin{itemize}
    \item \textbf{Recommendation-Specific Membership Characterization.}
    We characterize membership inference at the user, interaction, and social-relation levels, clarifying their objectives and distinctions from related privacy attacks. We also discuss membership inference in LLM-based RecSys, where sensitive information may be incorporated into inference prompts rather than model training data.

    \item \textbf{Multi-Dimensional Taxonomy and Comparative Analysis.}
    We systematically categorize existing MIAs according to their membership targets, adversarial knowledge, and recommendation architectures. We analyze the underlying attack mechanisms, key contributions, empirical findings, and limitations of representative approaches, highlighting how recommendation-specific characteristics influence membership leakage.

    \item \textbf{Evaluation and Defense Mechanisms.}
    We summarize commonly used datasets, recommendation models, and evaluation metrics for RecSys MIAs, with particular emphasis on attack effectiveness at low false-positive rates. We further review existing defense mechanisms, including differential privacy, regularization, machine unlearning, and privacy risk estimation, and discuss their effectiveness, privacy guarantees, and recommendation utility trade-offs.

    \item \textbf{Emerging Challenges and Future Research Directions.}
    We identify open research problems associated with federated, cross-domain, multimodal, and LLM-based recommendation architectures. In particular, we discuss evolving membership targets, realistic adversarial assumptions, and privacy risks arising from different stages of LLM-based personalization, providing research directions for advancing membership privacy in RecSys.
\end{itemize}

To support continuing research, we also maintain an open-source repository\footnote{\url{https://github.com/Richardwarriors/Membership-Inference-Attacks-on-Recommendation-System}} containing relevant publications and available implementations, which can be updated as new studies emerge.

The remainder of this paper is organized as follows. Section~\ref{sec:Preliminary} introduces the foundations of recommender systems and membership inference attacks. Section~\ref{sec:MIAsRecSys} presents the taxonomy and technical analysis of existing MIAs against RecSys. Section~\ref{sec:Defense} reviews privacy defense mechanisms, followed by an evaluation of commonly used datasets, recommendation models, and attack metrics. Section~\ref{sec:future and direction} discusses open challenges and future research opportunities, and Section~\ref{sec:conclusion} concludes this survey.

\section{Preliminaries}
\label{sec:Preliminary}

\subsection{Recommender Systems (RecSys)}

Recommender systems (RecSys) aim to predict user preferences and provide personalized item recommendations based on historical interactions and contextual information. They have been widely deployed in various applications, including job matching~\cite{10.1145/3543507.3583355, 10.1145/3209978.3210025,10.1145/3626772.3657776}, e-commerce~\cite{chen2023knowledgegraphcompletionmodels,10.1145/3219819.3219823,1167344}, and online entertainment~\cite{gao2023chatrecinteractiveexplainablellmsaugmented,10.1145/2959100.2959190,10.1145/2843948}. By modeling user--item interactions, RecSys learn user preferences and identify relevant items from large candidate collections~\cite{rendle2012bprbayesianpersonalizedranking,he2017neural,he2020lightgcn}.

Formally, let $\mathcal{U}$ and $\mathcal{I}$ denote the sets of users and items, respectively. A recommendation dataset is represented as
\[
\mathcal{D} = \{(u_j,i_j,y_j,c_j)\}_{j=1}^{N},
\]
where $u_j \in \mathcal{U}$, $i_j \in \mathcal{I}$, $y_j$ represents the observed user--item feedback (e.g., clicks, purchases, or ratings), and $c_j$ denotes optional contextual information, such as timestamps or interaction conditions. The historical interactions of user $u$ are denoted by $\mathcal{H}_u$. Given a training dataset $\mathcal{D}_{\mathrm{train}}$, a recommender model learns parameters $\theta$ by optimizing a recommendation objective:
\[
\theta^* =
\arg\min_{\theta}
\mathcal{L}_{\mathrm{rec}}
(\theta;\mathcal{D}_{\mathrm{train}})
+
\lambda\Omega(\theta),
\]
where $\mathcal{L}_{\mathrm{rec}}$ denotes the recommendation loss, $\Omega(\theta)$ is an optional regularization term, and $\lambda$ controls its contribution. The recommendation loss may be defined using pointwise prediction\cite{10.1145/3543507.3583355,10.1145/3298689.3347037}, pairwise ranking\cite{10.1145/3460231.3474274,10.1145/3109859.3109906}, or other learning objectives\cite{10.1145/3604915.3610236}, depending on the model architecture.

For conventional RecSys, the trained model estimates a preference score $s_{\theta}(u,i\mid\mathcal{H}_u,c)$ for candidate item $i$ and target user $u$. The system then generates a ranked list of $K$ recommended items:
\[
R_{\theta}(u,c)
=
\operatorname{TopK}_{i\in\mathcal{I}_{u}^{\mathrm{cand}}}
s_{\theta}(u,i\mid\mathcal{H}_u,c),
\]
where $\mathcal{I}_{u}^{\mathrm{cand}}$ denotes the candidate item set and $c$ represents the current recommendation context. Different RecSys architectures adopt different methods to learn user preferences and estimate item relevance. Early approaches, such as matrix factorization (MF), represent users and items in low-dimensional latent spaces to capture collaborative signals. More recent models, including NeuMF~\cite{he2017neural}, LightGCN~\cite{he2020lightgcn}, and SimpleX~\cite{mao2023simplexsimplestrongbaseline}, employ neural networks and graph-based representations\cite{Wang_2019,10.1145/3640457.3688064} to capture complex user--item relationships.

With the development of large language models (LLMs), recommendation has further evolved toward language-based modeling and generation. Representative approaches, including P5~\cite{10.1145/3523227.3546767}, M6-Rec~\cite{cui2022m6recgenerativepretrainedlanguage}, and TALLRec~\cite{Bao_2023}, incorporate textual user histories, item descriptions, and task instructions into recommendation. While some LLM-based RecSys retain conventional scoring and ranking pipelines, generative approaches can directly produce recommendations conditioned on an input prompt. Specifically, given a prompt $P_u$ containing user preferences, historical interactions, and task instructions, an LLM parameterized by $\phi$ generates a recommendation response:
\[
Y_u \sim p_{\phi}(\cdot\mid P_u),
\]
where $Y_u$ may contain recommended items, ranked lists, or natural-language explanations. Unlike conventional RecSys, where user information is primarily incorporated through model training, LLM-based RecSys may additionally utilize private user information dynamically through inference-time prompts.

These differences are particularly relevant to membership inference attacks (MIAs). In conventional RecSys, membership inference generally aims to determine whether a user or a specific interaction was included in $\mathcal{D}_{\mathrm{train}}$. For LLM-based RecSys, membership inference may additionally target whether private user information is included in the inference context $P_u$, rather than the model's training data. Therefore, distinguishing training-data membership from inference-context membership is important for understanding privacy risks across different recommendation architectures.

\subsection{Membership Inference Attacks (MIAs)}
Given a trained target model $f$ and a target record $z$, an adversary aims to determine whether $z$ was included in the training dataset of $f$. This can be formulated as a binary hypothesis testing problem:
\[
H_0:\; z \notin \mathcal{D} \quad\text{vs.}\quad H_1:\; z \in \mathcal{D},
\]
where $\mathcal{D}$ denotes the training set. The adversary computes a decision statistic based on information obtained from the target model---such as the output confidence\cite{shokri2017membership,salem2018mlleaksmodeldataindependent,song2021systematic}, likelihood ratio\cite{carlini2022lira, pmlr-v235-zarifzadeh24a}, or training loss trajectory\cite{10.1145/3658644.3690335,10.1145/3658644.3690335}---and compares it against a threshold to decide between the two hypotheses. Intuitively, records that the model has previously seen (members) tend to produce similar output characteristics than unseen records (non-members), due to phenomena such as model overfitting or memorization. Therefore, a membership inference attack (MIA) can be interpreted as a model-based distinguishing attack that exploits these behavioral discrepancies to infer the presence of specific records in the training data.

\section{Membership Inference Attacks on Recommendation Systems}
\label{sec:MIAsRecSys}
In this section, we first define membership inference attacks (MIAs) in recommender systems (RecSys) and distinguish them from related privacy inference tasks. We then discuss adversarial knowledge, attack taxonomies, membership targets, and recommendation architectures, highlighting the characteristics that shape MIAs in RecSys.

\subsection{Definition of MIAs on RecSys}
\begin{figure}[htbp]
  \centering
  \includegraphics[width=1.0\textwidth]{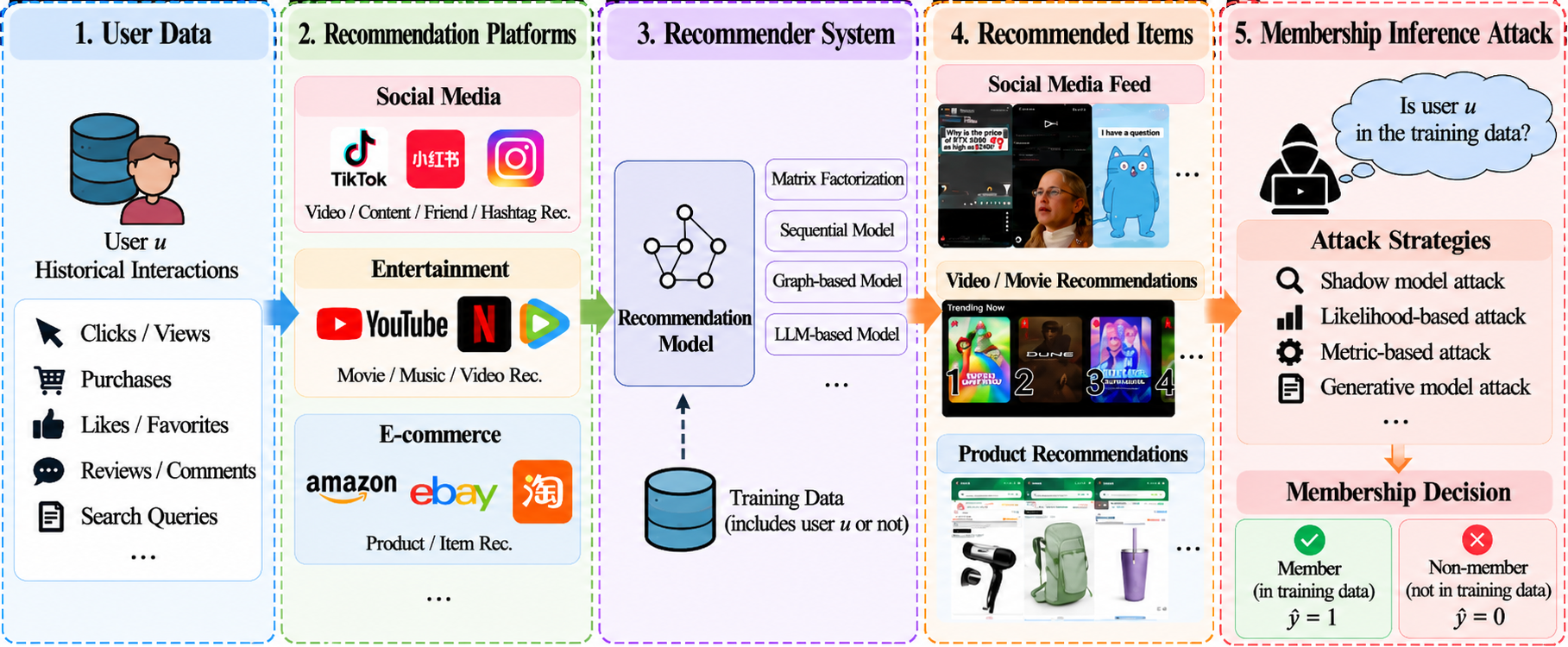}
  \caption{The Framework of MIA in Recommendation System.}
  \label{fig:framework}
\end{figure}

To illustrate MIAs in RecSys, Figure~\ref{fig:framework} presents a typical attack framework in which an adversary observes information exposed by a trained recommender and attempts to determine whether a target contributed to its training data. Unlike inference of a user's private attributes or relationships, the defining objective of a training-data MIA is to infer \emph{data inclusion}. Let $\mathcal{D}_{\mathrm{train}}$ denote the interaction records used to train the target recommender, and let $z_{ui}$ denote a specific user--item interaction record. We distinguish the following membership indicators:

\begin{equation}
\label{eq:membership_granularity}
\begin{aligned}
m_u
&= \mathbb{I}\!\left[
u \in \mathcal{U}_{\mathrm{train}}
\right], \\[6pt]
m_{ui}
&= \mathbb{I}\!\left[
z_{ui} \in \mathcal{D}_{\mathrm{train}}
\right], \\[6pt]
m^S_{uv}
&= \mathbb{I}\!\left[
(u,v) \in \mathcal{E}^{\mathrm{train}}_S
\right].
\end{aligned}
\end{equation}

where $\mathbb{I}[\cdot]$ is an indicator function and $\mathcal{E}^{\mathrm{train}}_S$ is the set of social relations used to train a social recommender. These indicators correspond to \emph{user-level}, \emph{interaction-level}, and \emph{social-relation-level} membership, respectively. A membership attack estimates one of these indicators from the information available under its threat model.

\subsection{Threat Models}
The amount and type of information available to an attacker critically determine the feasibility and strength of membership inference attacks (MIAs) against recommendation systems (RecSys). In this section, we first formalize the attacker's knowledge and then describe black-box and white-box MIAs in the RecSys setting. Broadly speaking, two classes of background knowledge are most relevant: (i) knowledge about the training data distribution, and (ii) knowledge about the target model.

\textbf{Knowledge about the training data distribution.} It refers to the attacker's understanding of the distribution from which the model was trained. Many MIA formulations assume that the adversary can access or synthesize a shadow dataset drawn from the same distribution as the target training data. This assumption is justified in practice: when distributional statistics are available, the shadow dataset can be generated via statistics-based synthesis\cite{salem2018mlleaksmodeldataindependent}, whereas in other cases, one may employ model-based synthesis techniques to approximate the underlying distribution \cite{shokri17}. For nontrivial evaluation, it is typically assumed that the shadow dataset is disjoint from the target training set.

\textbf{Knowledge about the target model} It captures information about how the RecSys is trained and parameterized, including posterior probability, and hyper-parameters (e.g., negative-sampling ratio).

RecSys raises an additional practical requirement not commonly emphasized in other domains: besides a shadow dataset, many RecSys attacks assume access to an \emph{item-embedding generating} (IEG) dataset that is disjoint from both the target and shadow datasets. The IEG dataset is used to produce item embeddings for the full item catalog, a step necessary because items are typically known a priori, even when user interaction traces are private. Requiring a separate IEG dataset is a mild and realistic assumption---platforms and public catalogs make item descriptions readily available---yet it materially affects attack design and transferability. Based on adversarial knowledge, we can characterize the dangerous levels of existing attacks.

\textbf{White-box Attacks.} Under this setting, an attacker can get some inaccessible information and use it to attack a target RecSys model. The information includes the posterior probability of the item and the learned parameters of the target model.

\textbf{Black-box Attacks.} In this setting, an adversary has only black-box access to the target recommendation model. The attacker may possess background knowledge about the training data distribution and users' historical interactions, but cannot directly access the model's internal parameters or intermediate representations. Instead, membership inference must rely on observable outputs, such as the top-\(K\) recommendation lists returned by the target system.

\begin{figure}[htbp]
  \centering
  \includegraphics[width=0.8\textwidth]{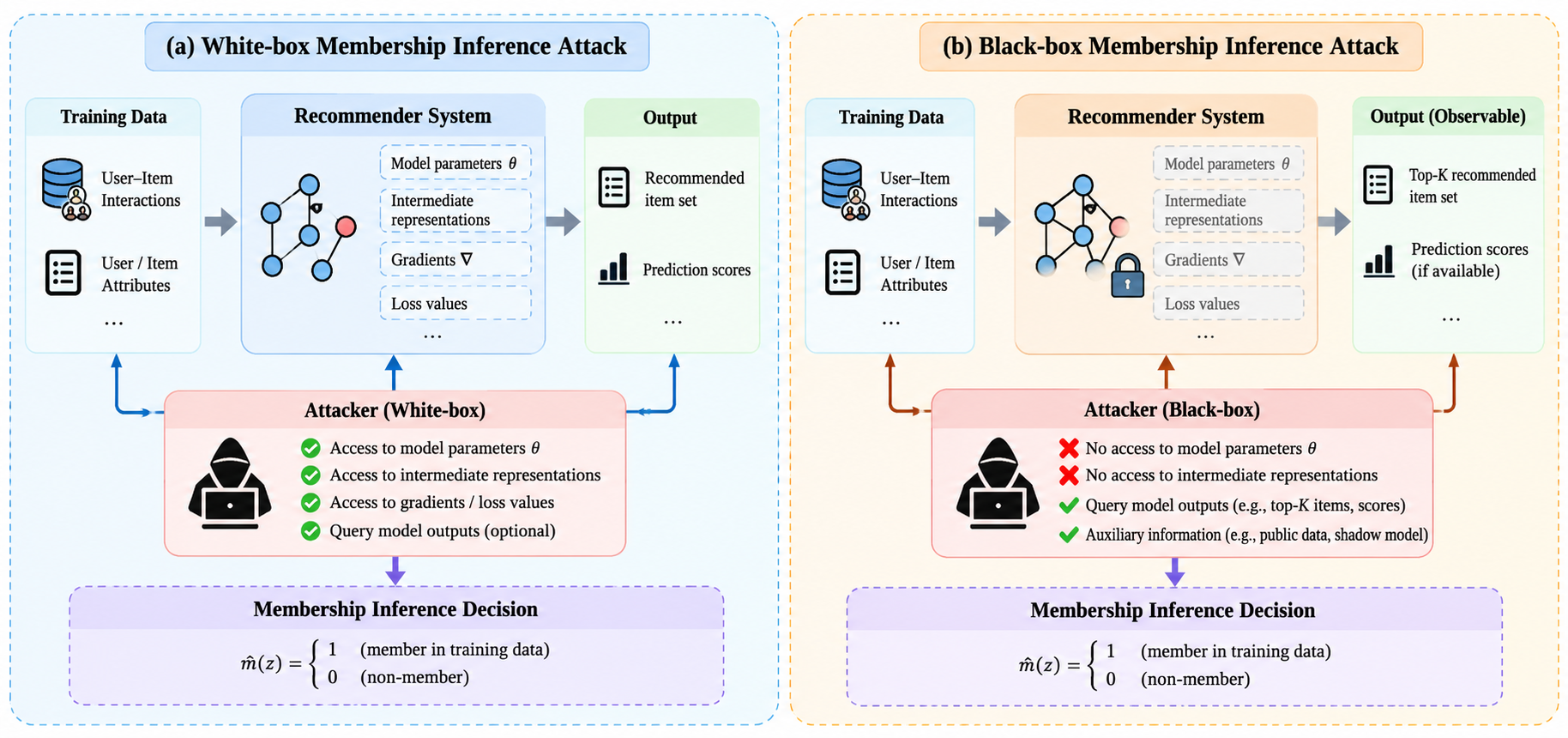}
  \caption{The Overview of white-box and black-box MIA in Recommendation System.}
  \label{fig:white_black}
\end{figure}

Figure~\ref{fig:white_black} illustrates the conceptual distinction between white-box and black-box MIAs on a target RecSys model. A green tick (\textcolor{green!60!black}{\ding{51}}) indicates available information, while a red cross (\textcolor{red!70!black}{\ding{55}}) denotes inaccessible information. As shown, the white-box adversary possesses access to detailed model internals, including learned parameters and posterior probabilities, whereas the black-box adversary is restricted to external query responses and user interaction data.

In the real world, obtaining access to model parameters or training configurations is exceedingly rare. Consequently, most recent research on MIAs in RecSys focuses on the black-box setting, which better reflects realistic deployment scenarios. The white-box setting, by contrast, is primarily explored in the context of Federated RecSys~\cite{yuan2023interaction}, where parameter updates may be partially exposed to participating clients. In traditional centralized RecSys models, white-box analyses are often used from a developer's standpoint to assess privacy risk scoring~\cite{10.1145/3705328.3748052} and proactively safeguard users identified as privacy-sensitive.

Although black-box attackers operate with significantly less information, the fact that effective black-box MIAs can still succeed underscores a critical vulnerability in modern RecSys. Demonstrating attack success under such limited information further emphasizes the urgent need for robust privacy-preserving mechanisms in RecSys.

\subsection{Taxonomies of Membership Inference Attacks on Recommendation Systems}

To systematically organize existing MIAs against RecSys, we establish a multi-dimensional taxonomy, summarized in Table~\ref{tab:mia_taxonomy}. Specifically, we categorize existing studies according to five complementary dimensions: (i) membership target, including user-level, interaction-level, and social-relation-level inference; (ii) recommender architecture, covering different recommendation models and their associated membership leakage mechanisms; (iii) training paradigm, distinguishing centralized and federated recommendation; (iv) adversarial access, including white-box and black-box settings; and (v) membership source, distinguishing training-data membership from inference-context membership in LLM-based RecSys.

These dimensions are not mutually exclusive, as a particular attack may target multiple recommendation architectures or membership granularities. Furthermore, distinguishing membership targets from membership sources is especially important for LLM-based RecSys, where private user information may be incorporated into either model training or inference-time personalization. This taxonomy provides a unified framework for comparing the objectives, assumptions, and characteristics of existing RecSys MIAs.

\begin{table*}[t]
\centering
\caption{A multi-dimensional taxonomy of membership
inference attacks against recommender systems.}
\label{tab:mia_taxonomy}

\footnotesize
\setlength{\tabcolsep}{5pt}
\renewcommand{\arraystretch}{1.12}

\begin{tabularx}{\textwidth}{
    @{}
    p{0.12\textwidth}
    p{0.20\textwidth}
    >{\raggedright\arraybackslash}X
    >{\raggedright\arraybackslash}p{0.28\textwidth}
    @{}
}
\toprule
\textbf{Dimension} &
\textbf{Category} &
\textbf{Key Characteristic} &
\textbf{Representative Studies} \\
\midrule

% I. Membership Target
\multirow{3}{*}{\textbf{Target}}
& User-level
& User training participation
& \cite{zhang2021membership,wang2022debiasing} \\

& Interaction-level
& User--item training records
& \cite{zhong2024interaction,10.1145/3705328.3748052} \\

& Social-relation-level
& Private user--user ties
& \cite{10.1145/3726302.3730086} \\

\midrule

% II. Recommender Architecture
\multirow{6}{*}{\textbf{Model}}
& MF-based
& Latent-factor differences
& \cite{zhang2021membership,10.1145/3705328.3748052} \\

& Sequential
& Temporal and ranking signals
& \cite{zhu2023membership,chi2024shadowfreemembershipinferenceattacks} \\

& Graph/KG-based
& Graph and KG representations
& \cite{zhong2024interaction,10.1145/3726302.3730086} \\

& Cross-domain
& Cross-domain preference transfer
& \cite{11036029} \\

& Hybrid
& Reference recommendations
& \cite{11036029} \\

& LLM-based
& In-context behavior / distillation
& \cite{10.1145/3773078.3831733,10.1609/aaai.v40i27.39449} \\

\midrule

% III. Training Paradigm
\multirow{2}{*}{\textbf{Training}}
& Centralized
& Centrally trained models
& \cite{zhang2021membership,zhong2024interaction} \\

& Federated
& Client updates and shared parameters
& \cite{yuan2023interaction,11372209} \\

\midrule

% IV. Adversarial Access
\multirow{2}{*}{\textbf{Access}}
& White-box
& Model parameters and updates
& \cite{yuan2023interaction,10.1145/3705328.3748052} \\

& Black-box
& Recommendation outputs
& \cite{chi2024shadowfreemembershipinferenceattacks,zhu2023membership} \\

\midrule

% V. Membership Basis
\multirow{2}{*}{\textbf{Basis}}
& Training data
& Training-set inclusion
& \cite{10.1145/3705328.3748052,10.1609/aaai.v40i27.39449} \\

& Inference context
& Prompt/context inclusion
& \cite{10.1145/3773078.3831733} \\

\bottomrule
\end{tabularx}
\end{table*}

\subsection{Targets of Membership Inference Attacks on Recommendation Systems}
\label{sec:rec-mia-targets}
Recommendation systems are inherently complex information-fusion systems that integrate diverse sources of user and item data. As a result, they encapsulate multiple levels of privacy-sensitive information from the adversary's perspective. The training dataset for a RecSys typically contains rich, heterogeneous information, including user attributes, behavioral histories, and social relationships. From these different aspects, adversaries can launch various membership inference attacks (MIAs) targeting distinct forms of private information. From the user level, the adversary observes the behavior of the target RecSys on member records (i.e., data points used during training) versus non-member records (i.e., data points unseen during training) to infer whether a specific user was included in the training dataset. From the interaction level, the focus shifts to user behaviors and preferences, which tend to be more sensitive. Here, the attacker aims to infer whether a particular user--item interaction (e.g., a purchase, click, or rating) was part of the training data, thereby revealing a user's interests, preferences, or even daily habits. Finally, at the social level, which arises in social or graph-based recommendation systems, the adversary attempts to exploit the collaborative filtering or graph embeddings associated with recommended items to infer hidden social ties---such as whether two users are connected in the underlying social network.

Overall, these three attack granularities---\textbf{user-level}, \textbf{interaction-level}, and \textbf{social-level} MIAs---represent the principal dimensions along which privacy risks manifest in modern RecSys.

Although these categories may resemble attribute inference, feature inference, and link inference, their objectives differ. \emph{Attribute inference} aims to recover an unknown user attribute (e.g., demographic information), while \emph{feature inference} estimates an unobserved feature or field of a data record, irrespective of whether the record was used for model training. By contrast, an interaction-level MIA tests whether a specified interaction record belongs to $\mathcal{D}_{\mathrm{train}}$, rather than merely predicting whether the user has a preference for the item. Likewise, \emph{link inference} determines whether an edge exists in an underlying interaction or social graph, whereas social-relation-level training-data membership tests whether an edge was included in $\mathcal{E}^{\mathrm{train}}_S$. These two tasks coincide only under a data construction protocol in which the target graph is identical to the graph used for training. Therefore, the interpretation of a reported attack depends on its ground-truth membership labels and the relationship between the observed graph and the training graph. In the following subsections, we retain the attack terminology used in the original studies while making these distinctions explicit.

\subsubsection{User-Level MIAs}
Zhang et al.\cite{zhang2021membership} were the first to propose the user-level membership inference attack in RecSys based on item embedding differences, aiming to infer whether a user's data was included in the training dataset of the RecSys model. Under the black-box setting, the attacker can only observe the output recommendation list. The item-embedding-based method assumes that if a user's data is included in the training set, the recommended items should be similar to the items the user has interacted with. In contrast, for non-member users, the system lacks knowledge of their preferences and therefore cannot generate personalized recommendations. In the original work, the top popular items were used as the recommendations for non-members. To construct attack features, the embeddings of the interacted items were averaged, and those of the recommended items were averaged; the difference between the two was then used as the membership feature. A two-layer multilayer perceptron (MLP) was trained as the attack model to distinguish members from non-members.

However, this approach is unstable when the target and shadow datasets differ. The reason is that the method implicitly assumes that item embeddings are generated by a fixed collaborative filtering model. In practice, embeddings trained by different algorithms (e.g., BPR\cite{rendle2012bprbayesianpersonalizedranking} vs. NeuMF\cite{he2017neural}) can diverge in the latent space even when trained on the same dataset. Furthermore, the item embeddings generated by the target and shadow models may also exhibit discrepancies, which degrade attack performance. To address this issue, Wang et al. \cite{wang2022debiasing} proposed a debiased learning MIA (DL-MIA) framework for RecSys that mitigates the embedding bias between the target and shadow models. For DL-MIA, to overcome the limitations of Item Difference MIAs (ID-MIA), which suffer from (i) a training-data bias---distributional gaps between shadow and target recommenders that make attack samples generated by the shadow model poorly transferable---and (ii) an estimation bias---since the attacker cannot observe hidden states (user/item embeddings), externally constructed difference vectors are noisy---the DL-MIA framework explicitly debiases learning within the difference-vector paradigm. Concretely, it (1) builds difference vectors from each user's history and the system's recommendation list; (2) employs a variational auto-encoder--based disentangled encoder to separate recommender-invariant from recommender-specific features, narrowing the shadow--target gap and mitigating training-data bias; (3) learns a truth-level score per difference vector as a sample weight to discount poorly estimated features, thereby mitigating estimation bias; and (4) trains a member/non-member classifier (MLP) on the disentangled, reweighted representations, optimized with an alternating training procedure. Empirically, DL-MIA simultaneously reduces both biases and achieves state-of-the-art attack performance across general and sequential RecSys.

Although subsequent experiments demonstrated that DL-MIA is more accurate and stable than the ID-MIA under target--shadow mismatches, its effectiveness remains sensitive to the information exposed to the adversary---most notably the dimensionality of item embeddings and the length of the top \textendash $K$ recommendation list. Empirically, performance tends to improve with larger embedding dimensions and larger $K$, whereas short lists (e.g., $K \le 10$) often provide a weak signal in practice. Moreover, obtaining the training dataset distribution is unrealistic. Chi et~al\cite{chi2024shadowfreemembershipinferenceattacks}. proposed the shadow-free MIA method to infer the membership without training the shadow model, compared with the previous methods\cite{zhang2021membership,wang2022debiasing,zhu2023membership}. The intuition of the SF-MIA is to compare the recommended items with general popular items (which can be obtained by creating an empty account). If the recommended items align more closely with popular items, then it's likely not a member. But if the recommended items have a higher similarity to the user's historical interactions, then it'll be classified as a member. The experiment results show that this shadow-free approach not only significantly reduces computational cost for MIAs but also reaches a comparable performance to previous shadow-training approaches. At the same time, since the shadow model does not need to be trained, this method reduces the attack's time cost. However, the common assumption that non\textendash member recommendations align with a globally popular item set is brittle in the real world. The current sequential RecSys can provide personalized recommendations for both member and non-member users. To address the member/non-member mode gap, Zhu et al.\cite{zhu2023membership} propose the Model Extraction based MIA (ME-MIA) for sequential recommenders, which first extracts a surrogate model via black\textendash box queries that align its recommended item set and rank order with the target using generic list\textendash consistency objectives, and then performs membership inference using the surrogate's richer signals (scores, ranks, similarities). ME-MIA targets sequential recommenders in a black-box setting by first extracting a surrogate model that imitates the target's top \textendash $K$ items and their rank order using two generic objectives---ranking consistency and positive-item consistency---derived solely from recommendation lists. Because the surrogate's training does not reveal membership labels, a shadow model (trained with the same objective) is used to construct labeled data; rich signals from the surrogate (scores, ranks, similarities) then feed a binary classifier for membership inference. To reduce data demands, ME-MIA offers (i) a data-efficient variant that augments sequences by replacing actions with nearest-neighbor items in the surrogate's embedding space, and (ii) a data-free variant that synthesizes sequences autoregressively and queries the target for soft labels, following data-free model-extraction practice. These steps yield effective, transferable attacks in black-box sequential settings. While ME-MIA demonstrates that membership inference is possible even in data-free settings, this comes at the cost of effectiveness; moreover, the role of item scoring in shaping attack success remains underexplored, and it is unclear whether ME-MIA readily generalizes to non-sequential RecSys. Despite recent progress at the user level, open challenges persist: how to design more efficient attacks, reduce confounding factors that obscure the signal (e.g., embedding dimensionality, top \textendash $K$, popularity effects), and accommodate the inherent heterogeneity of recommender systems with model\textendash agnostic methods. Beyond traditional RecSys MIAs, He et~al.\cite{10.1145/3773078.3831733} present the first user\textendash level attacks for \emph{LLM\textendash based} recommender systems, proposing two attacks---\emph{Contextual Item Memorization Attack}, \emph{Intertia Attack} --- that exploit LLMs' short-term memorization, memorization confidence; evaluated across multiple popular LLMs, these attacks demonstrate clear effectiveness in revealing whether specific users were included in in\textendash context prompts.

\subsubsection{Interaction-level MIAs}
RecSys poses privacy risks beyond the user level: in addition to concerns about whether a user is included in the training set, one may also be interested in the presence or absence of specific user--item interactions, referred to as \emph{interaction-level privacy}. In this part, we focus on centralized RecSys, while the discussion of Federated RecSys will be presented in Section~\ref{sec:fed-mias}. Compared to user-level privacy studies, research on interaction-level privacy remains relatively limited. Although the privacy risks associated with interaction-level breaches are often more severe, designing effective MIAs at this level is considerably more challenging, primarily because existing embedding-difference-based methods are not directly applicable. To address the challenge of interaction-level MIA on RecSys, Zhong et~al.\cite{zhong2024interaction} firstly proposed interaction-level MIA called MINER on a knowledge-graph (KG)--based RecSys, which is a framework that learns to infer whether a specific user-item interaction was included in the training data by leveraging knowledge-enhanced embeddings and a bilateral-branch attack model. The intuition of MINER is measuring the distance similarity metric between the interacted item and recommended item to infer the member and non-member data. For ranked recommendation lists, MINER computes a discounted similarity score (DS) that logarithmically weights higher-ranked items more heavily:
\[
DS(i, i') = \frac{d(\mathbf{e}_i, \mathbf{e}_{i'})}{\log_2(r_{i'} + 1)},
\]
where $d(\cdot)$ denotes a distance function and $r_{i'}$ is the ranking of item $i'$ in the top-$k$ recommendation list. Multiple distance metrics (L1, L2, cosine, Bray--Curtis) are used, and their concatenation forms the feature vector $\mathbf{x}$ for each user--item pair.

Considering the distribution of the item, the author thinks the personalized interacted item (non-popular item) contains more sensitive information. For example, consider a healthcare RecSys. Determining whether a patient has been treated (\textit{}``interacted'') with HIV (a rare disease) carries greater sensitivity than discerning whether the user has been treated with flu, a common disease. MINER introduces a bilateral-branch attack model with two sub-networks: a \textit{main branch} trained on the original long-tailed distribution and a \textit{regularizer branch} trained on a re-balanced distribution, which enables each branch to learn complementary knowledge from head and tail interactions, respectively. Through this bilateral learning strategy, MINER effectively mitigates the influence of long-tailed distributions and achieves high attack accuracy on both head and tail interactions.

However, while MINER targets the long-tail, it shows limited effectiveness in the low-FPR regime and its reliance on KGs constrains applicability to other RecSys architectures. To address the model-agnostic constraint, He et al.~\cite{10.1145/3705328.3748052} adapt the Likelihood Ratio Attack (LiRA)~\cite{carlini2022lira} to the recommender systems setting and propose \textit{RecLiRA}. The core idea is that the posterior confidence distributions of member (IN) and non-member (OUT) samples exhibit measurable differences, enabling privacy risk to be quantified through their statistical separability. For each shadow model, RecLiRA collects confidence scores for IN and OUT samples and models them as Gaussian distributions. A statistic $q = |2p - 1|$ and its logit transformation $\phi(q) = \log\!\left(\frac{q}{1-q}\right)$ are then used to better distinguish between the two distributions. RecLiRA is versatile and can be applied to both \textit{interaction-level} and \textit{user-level} membership inference attacks.

For interaction-level attacks such as MINER and RecLiRA, the relevant membership question is whether the target user--item record was used during the recommender's training, not simply whether the user ever interacted with the item. The latter constitutes behavioral or interaction-existence inference unless the evaluation protocol explicitly equates observed interactions with training records. 

\subsubsection{Social-Level MIAs}
Unlike user-level and interaction-level MIAs, which only require knowing whether a user or a user's interactions was included in the training dataset, Zhao et~al.\cite{10.1145/3726302.3730086} introduces a social-level Membership Inference Attack (SMIA) framework, which moves beyond traditional user- and interaction-level attacks to infer whether a social link exists between two users in a social RecSys. The framework targets the inference of whether a user pair \((u_1,u_2)\) has a social relation in the social graph \(G_S\) of a social RecSys. The intuition is that users with social ties tend to have higher similarity in recommendation lists or embedding space (i.e., social homophily), allowing inference of hidden relationships. The attacker has black-box access to the target RecSys model \(M_{\text{target}}\) (i.e., only recommendation results), and optionally a shadow social graph \(G'_S\). The adversary firstly collects the recommendation outputs (top-\(k\) lists) from \(M_{\text{target}}\) for target users and forms a \emph{shadow interaction graph} \(G'_A\). This graph approximates how users are connected via item recommendations. Then, using \(G'_S\) and \(G'_A\), the attacker trains a dual-branch model:
\begin{itemize}
  \item \emph{Shadow Social Preference Learner:} takes the social graph \(G'_S\) and computes user representations via a GCN, aiming to capture social influence and homophily patterns (embeddings \(E^S\)).
  \item \emph{Shadow Behavioral Preference Learner:} uses the interaction graph \(G'_A\) to extract user behaviors independently (embeddings \(E^B\)).
\end{itemize}
The embeddings from both branches are then aggregated (e.g., concatenation, attention) to form a combined pair-feature representation for any user pair. Finally, the attacker trains a binary classifier to predict \(\hat y_{u_1,u_2}\). If a social relation exists between \(u_1\) and \(u_2\), \(\hat y_{u_1,u_2} = 1\); otherwise, \(\hat y_{u_1,u_2} = 0\). In short, SMIA reveals a previously underexplored privacy dimension, social-level inference in RecSys, by combining recommendation output analysis, shadow-model learning, and user-pair classification to infer hidden social ties.

The terminology of social-level MIA warrants an additional distinction from conventional link inference. Zhao et al.~\cite{10.1145/3726302.3730086} use the term \emph{social relation-level MIA} to describe inference of private user--user relationships from recommendation outputs. In a setting where the target relationship belongs to the private social graph used to train the recommender, this objective can be viewed as edge-level training-data membership. However, predicting the existence of a social tie does not, by itself, establish that the tie was present in the training graph. If training edges are a subset of the underlying social network, social-link existence and social-edge membership have different ground-truth labels. Accordingly, we retain SMIA under the social-level privacy category while explicitly recognizing its overlap with link inference and the need to state the underlying training-graph assumptions.

\subsection{Membership Inference Attacks on Different Recommendation Systems}

\subsubsection{Matrix-Factorization-Based RecSys}

Matrix factorization-based RecSys (MF-RecSys) represent
users and items as low-dimensional latent vectors and
predict user preferences through their interactions,
typically using inner products to capture collaborative
patterns in the user--item matrix. MF-based approaches
remain fundamental baselines for explicit- and
implicit-feedback recommendation, including
BPR-based ranking~\cite{rendle2012bprbayesianpersonalizedranking}
and neural extensions such as
NeuMF~\cite{he2017neural}. Early MIAs against RecSys primarily exploited differences between the recommendation behaviors of member and non-member users. Zhang et al.~\cite{zhang2021membership} introduced a user-level MIA that constructs membership features by comparing the embeddings of a user's historically interacted items with those of the recommended items. Building upon this approach, Wang et al.~\cite{wang2022debiasing} proposed DL-MIA to mitigate distributional discrepancies between target and shadow recommenders through debiased representation learning, improving attack robustness across different recommendation models. Moving beyond embedding-difference-based attacks, He et al.~\cite{10.1145/3705328.3748052} proposed
RecLiRA, a likelihood-ratio-based framework that exploits differences between the confidence distributions of member and non-member records. RecLiRA supports both user- and interaction-level membership inference across multiple RecSys architectures, achieving strong detection performance at low false positive rates. Furthermore, the authors introduced a differential privacy inspired privacy score,
$\ln(\mathrm{TPR}/\mathrm{FPR})$, to quantify membership leakage at the interaction level and aggregate privacy risks to the user level~\cite{He_2025}.

Overall, these studies demonstrate that MF-based recommendation models can expose membership signals through both latent representation differences and prediction confidence distributions. However, their effectiveness may depend on the quality of auxiliary information and the consistency between target and shadow models.

\subsubsection{MIAs on Sequential RecSys}

Sequential RecSys model users' evolving preferences by capturing temporal dependencies in their historical interactions. Unlike conventional collaborative filtering approaches that primarily rely on unordered user--item interactions, sequential models leverage
interaction order to predict future user behavior. They have been widely applied to e-commerce
recommendation~\cite{8594844}, music streaming~\cite{10.1145/3488560.3498477}, and online content recommendation~\cite{tang2018personalized}. Representative architectures have evolved from recurrent models such as GRU4Rec~\cite{hidasi2015gru4rec} to self-attention-based models, including SASRec~\cite{8594844} and BERT4Rec~\cite{sun2019bert4rec}. While earlier user-level MIAs based on recommendation and interaction similarities can also be applied to sequential recommenders, recent studies have explored more practical adversarial settings. Zhu et al.~\cite{zhu2023membership} proposed ME-MIA, a black-box attack that extracts a surrogate sequential recommender by approximating the target model's ranking behavior. Membership inference is then performed using richer surrogate-model signals, including prediction scores, ranks, and item similarities. The framework further introduces
data-efficient and data-free variants to reduce dependence on auxiliary user interactions. To eliminate the computational overhead associated with training shadow models, Chi et
al.~\cite{chi2024shadowfreemembershipinferenceattacks} developed SF-MIA, a shadow-free approach that compares a user's recommendation list with both historically interacted items and globally popular items. The attack exploits differences in recommendation similarity to distinguish members from non-members, achieving competitive performance without requiring additional shadow-model training.

These developments highlight the feasibility of membership inference against sequential recommenders under increasingly restrictive adversarial assumptions. Nevertheless, the effectiveness of such attacks may remain sensitive to recommendation list length, popularity bias, and the availability of representative auxiliary data.

\subsubsection{MIAs on Graph-Based RecSys}
Graph-based RecSys model user--item interactions as graphs, where nodes represent users or items and edges correspond to interactions such as ratings, clicks, or purchases. By leveraging the rich relational structure inherent in these graphs, such models aim to learn high-quality embeddings that capture both connectivity patterns and higher-order collaborative signals. Graph-based RecSys have achieved remarkable success in modeling user preferences and item similarities through message-passing and neighborhood-aggregation mechanisms, as demonstrated in representative models such as LightGCN~\cite{he2020lightgcn}, CKE~\cite{zhang2016cke}, and KGAT~\cite{wang2019kgat}. These approaches extend traditional collaborative filtering by propagating information along user--item bipartite graphs to capture multi-hop dependencies.

Despite their effectiveness, the graph-based paradigm introduces new privacy risks. The learned embeddings inherently encode users' interaction behaviors and neighborhood relationships, which can be exploited by Membership Inference Attacks (MIAs). Yuan et al.~\cite{yuan2023interaction} first proposed an \emph{interaction-level} MIA on Federated RecSys, while Zhong et al.~\cite{zhong2024interaction} extended this idea to centralized RecSys. However, both of these studies focus solely on the interaction level. More recently, Zhao et al.~\cite{10.1145/3726302.3730086} discovered that edge connections between users can introduce a new form of sensitive information leakage, termed \emph{social-level privacy risk}. In this setting, the adversary's objective shifts from identifying individual user--item interactions to inferring social relationships among users---such as friendships, follow links, or communication patterns---on social media platforms. The intuition behind social-level MIAs lies in the distinguishable collaborative filtering patterns between members and non-members within the social graph, revealing privacy vulnerabilities beyond traditional interaction-level attacks. The relation to link inference depends on whether the edge-existence labels in the attack correspond to edges used for model training, as discussed in Section~\ref{sec:rec-mia-targets}.

\subsubsection{MIAs on Cross-Domain RecSys}

Cross-domain recommender systems (CDRS) aim to alleviate data sparsity and cold-start problems by transferring user preferences and item representations across different recommendation domains. However, the sharing and alignment of latent representations may introduce additional membership privacy risks, as information learned from one domain can influence recommendation outcomes in another. To investigate these vulnerabilities, Yang and Liu~\cite{11036029} proposed CDRMIA, a membership inference attack specifically designed for cross-domain recommender systems. Unlike conventional MIAs that focus on membership within a single recommendation domain, CDRMIA aims to infer a user's membership status in both source and target domains using target-domain recommendation outputs and partial behavioral information. The framework constructs cross-domain features based on embedding alignment, behavioral consistency, and prediction confidence, and employs a multi-task learning model to jointly predict membership states across domains. CDRMIA indicates that information transfer intended to improve recommendation quality may also expose membership signals across domain boundaries.Nevertheless, several challenges remain. The effectiveness of cross-domain MIAs may depend on the extent of user overlap, the strength of representation alignment, and the amount of auxiliary behavioral information available to the adversary. 

\subsubsection{MIAs on LLM-Based RecSys}

The integration of large language models (LLMs) into recommender systems introduces new membership privacy risks, as LLM-based RecSys can leverage user-specific information through both in-context learning (ICL) and model training. Unlike traditional RecSys, these systems incorporate textual user histories, item descriptions, and personalized prompts, creating additional attack surfaces for membership inference.

He et al.~\cite{10.1145/3773078.3831733} investigated MIAs against ICL-based RecSys, where users' historical interactions are incorporated into prompts to generate personalized recommendations. The authors proposed two attack strategies, \emph{ItemMem} and \emph{RecInertia}, which exploit membership-dependent differences in recommendation behavior to infer whether private user interactions are included in the inference context. More recently, Li et al.~\cite{10.1609/aaai.v40i27.39449} proposed a knowledge distillation-based MIA framework to address the high computational cost of conventional shadow-model-based attacks against LLM-based RecSys. Instead of training large shadow models from scratch, the framework constructs a reference model through knowledge distillation, employing distinct distillation strategies for member and non-member samples. It further integrates multiple membership-related features, including prediction confidence, entropy, loss, and hidden-layer representations, to train an attack classifier. Unlike ICL-based MIAs, which target information included in inference prompts, distillation-based MIAs focus on training-data membership. Nevertheless, their effectiveness depends on the availability of suitable auxiliary data and model-derived signals, highlighting the need for further evaluation under realistic adversarial constraints.

\subsection{Membership Inference Attacks on Federated Recommender Systems}
\label{sec:fed-mias}

Federated learning (FL) has emerged as an alternative to conventional centralized learning, where all training data are pooled to train a machine learning (ML) model. FL allows multiple parties to collaboratively train a shared model without directly exchanging their raw data, making it an attractive framework for privacy-sensitive applications~\cite{pmlr-v54-mcmahan17a,mcmahan2018learningdifferentiallyprivaterecurrent}. However, the success of membership inference attacks (MIAs) against FL has demonstrated that keeping data locally does not necessarily prevent privacy leakage. Melis et al.~\cite{melis2019exploiting} investigated privacy leakage in federated text classification models, where recurrent neural networks employ word-embedding layers to transform input sequences into low-dimensional representations. During training, the gradients of the embedding layer are sparse with respect to the input words: embeddings corresponding to words appearing in a local batch receive updates, while others remain unchanged. By observing these gradient updates, an adversary can infer information about the words present in private training data. Although MIAs have been extensively investigated in federated classification tasks, directly extending these approaches to federated recommender systems (Federated RecSys) remains challenging due to substantial differences in model architectures and training objectives.

\begin{figure}[htbp]
  \centering
  \includegraphics[width = 1.0\textwidth]{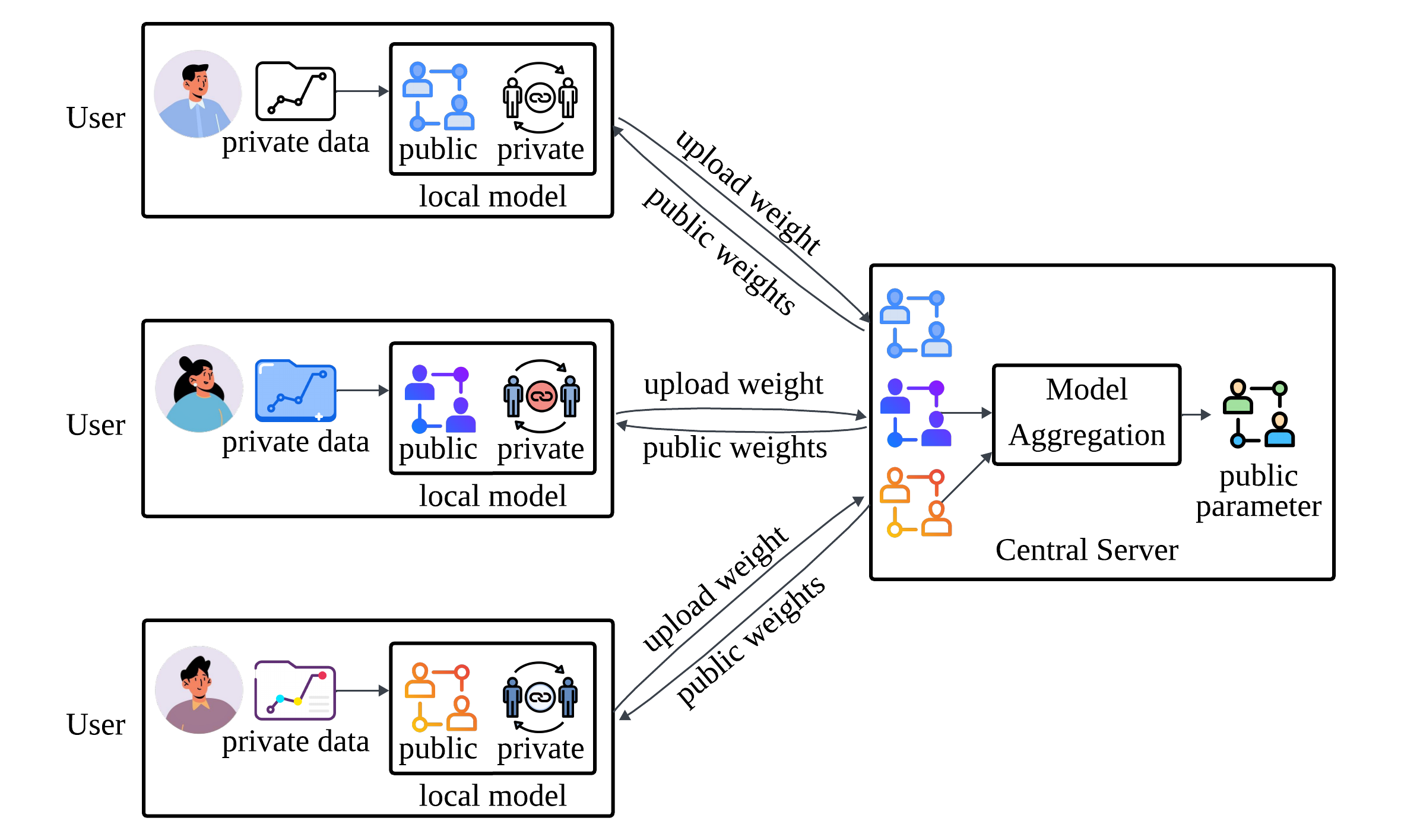}
  \caption{An overview of a federated recommender system.}
  \label{fig:Fed-Rec}
\end{figure}

A typical Federated RecSys framework is illustrated in Figure~\ref{fig:Fed-Rec}. Ammad et al.~\cite{ammaduddin2019federatedcollaborativefilteringprivacypreserving} proposed an early federated collaborative filtering framework, which inspired subsequent developments in decentralized recommendation. For example, FedFast~\cite{10.1145/3394486.3403176} aimed to accelerate the convergence of federated recommendation training, while Imran et al.~\cite{10.1145/3560486} and Wang et al.~\cite{wang2021fastadaptingprivacypreservingfederatedrecommender} focused on improving training efficiency and model adaptation. Despite these advances, several studies have questioned whether Federated RecSys provide sufficient privacy protection. Zhang et al.~\cite{zhang2023comprehensiveprivacyanalysisfederated} investigated privacy leakage in Federated RecSys, particularly the exposure of sensitive user attributes, and proposed corresponding protection mechanisms. Other studies~\cite{9170754,10.1145/3460231.3478855,10.1145/3548456} have also explored user information leakage and related defenses. Building upon these findings, Yuan et al.~\cite{yuan2023interaction} conducted an early systematic investigation of interaction-level MIAs against Federated RecSys, while more recent research has explored causal inference-based collaborative attacks~\cite{11372209}.

The main challenges of designing MIAs against Federated RecSys arise from two aspects. First, regarding the attack objective, MIAs in federated classification generally aim to determine whether a particular sample was used during training or whether it belonged to a specific client's local dataset. In Federated RecSys, the items involved in a client's local training process may be inferred by observing updates to shared item embeddings. However, these items include both positive and negative samples (i.e., interacted and non-interacted items), and only positive interactions directly reveal users' private preferences. Therefore, identifying the updated items alone is insufficient to infer their actual interaction labels. This illustrates the need to distinguish inference of a positive behavioral relation from inference that a specific record participated in local training; the two coincide only when the construction of the target membership labels makes that correspondence explicit. Second, from the attack implementation perspective, conventional federated MIAs often rely on auxiliary data or assumptions that may not hold in recommendation settings. Moreover, Federated RecSys commonly maintain private user embeddings while sharing item-related parameters, introducing architectural differences from conventional federated classification models.

Yuan et al.~\cite{yuan2023interaction} conducted the first systematic study of interaction-level membership inference attacks against Federated RecSys, proposing IFed-MIA. The framework considers a curious-but-honest central server attempting to infer a user's private positive interaction set $\mathcal{V}_i^+$. The server is assumed to observe the shared item parameters $\mathbf{V}_i^t$ uploaded by individual clients and relevant training hyperparameters, such as the learning rate and negative sampling ratio, without directly accessing private user embeddings or local interaction records. By analyzing which item embeddings are updated during local training, the server can identify candidate items involved in optimization but cannot directly determine whether these items correspond to positive or negative interactions. To infer the actual interaction label $r_{ij}$, IFed-MIA exploits an empirical distance principle involving three locally trained models: $M_i$, trained on the original dataset $\mathcal{D}_i$; $M'_i$, trained on the same dataset with a different initialization; and $M''_i$, trained on a modified dataset $\mathcal{D}_i^j$ where $r_{ij}$ is reversed. The resulting item embeddings are observed to satisfy
\[
\mathrm{dist}(\mathbf{v}_j,\mathbf{v}'_j)
<
\mathrm{dist}(\mathbf{v}_j,\mathbf{v}''_j).
\]
This observation suggests that reversing an interaction label may induce a larger embedding discrepancy than changing model initialization, providing a basis for distinguishing positive and negative interactions.

Since the server does not know the true interaction labels in $\mathcal{D}_i$, Yuan et al. construct a synthetic dataset $\mathcal{D}_i^{\mathrm{fake}}$ by assigning pseudo-labels to updated items according to the assumed negative sampling ratio (e.g., $1{:}4$). The attacker then trains a fake local model $M_i^{\mathrm{fake}}$ on $\mathcal{D}_i^{\mathrm{fake}}$ and compares its item embeddings with the observed client updates $\mathbf{V}_i^t$. Items with smaller embedding distances are identified as more likely to represent positive interactions, and the process iterates until the estimated proportion of positive samples is reached. The inference procedure can be executed asynchronously on the server side without interrupting the standard federated training process. Experimental results on Fed-NCF and Fed-LightGCN demonstrate that IFed-MIA achieves over 90\% interaction inference accuracy under the evaluated settings, revealing substantial privacy risks even without direct access to private user embeddings or raw interaction data. However, the attack relies on knowledge of training hyperparameters, particularly the negative sampling ratio, and its practical applicability under more restrictive adversarial assumptions remains uncertain.

More recently, Chen et al.~\cite{11372209} proposed CiCo-MIA, a causal inference-based collaborative membership inference framework designed to improve attack robustness and interpretability in Federated RecSys. Unlike IFed-MIA, which primarily exploits embedding distance patterns to identify private interactions, CiCo-MIA utilizes interaction data from multiple clients to construct a shadow recommender that approximates the target federated recommendation model. The framework further constructs a causal graph to characterize user--item dependencies and introduces a counterfactual reasoning-based user--item impact elimination module to reduce the influence of confounding interactions. By estimating the effects of user--item relationships through counterfactual reasoning, the attack aims to improve membership inference robustness and interpretability. In addition, a dynamic aggregation strategy is employed to balance attack effectiveness and recommendation performance. Furthermore, CiCo-MIA remains effective under local differential privacy (LDP) defenses, retaining approximately 30\% of its original performance in the reported experiments. These findings demonstrate that collaborative information and causal modeling can provide additional membership inference signals even when privacy-preserving mechanisms are applied. Nevertheless, the effectiveness of CiCo-MIA may depend on the availability of representative auxiliary interaction data and the reliability of the constructed causal relationships, motivating further investigation under heterogeneous client distributions and restricted adversarial access.

Overall, IFed-MIA and CiCo-MIA demonstrate that Federated RecSys remain vulnerable to membership leakage despite keeping raw user data decentralized. While IFed-MIA exploits shared item embedding updates, CiCo-MIA extends the attack methodology through collaborative shadow modeling and counterfactual reasoning. Their findings highlight the need for stronger defenses that protect private user interactions without substantially compromising recommendation utility.

\section{Defense Mechanisms}
\label{sec:Defense}
\begin{figure}[htbp]
  \centering
  \includegraphics[width = 1.0\textwidth]{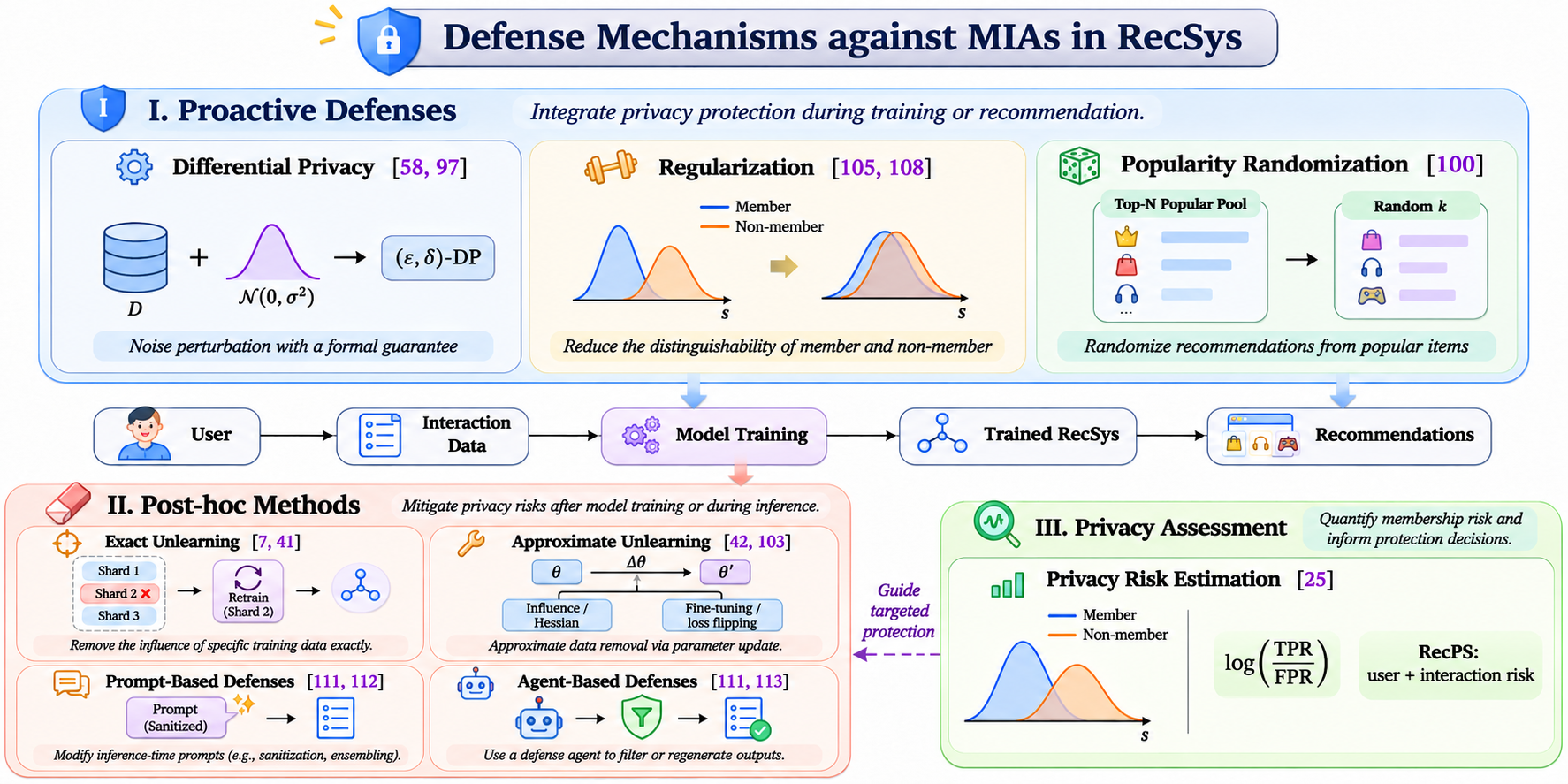}
  \caption{An overview of privacy protection against MIAs in recommender systems, including proactive defenses, post-hoc unlearning, and privacy risk estimation for guiding targeted protection.}
  \label{fig:Rec-Defense}
\end{figure}

Although research on membership inference attacks (MIAs) against RecSys has expanded, effective privacy protection mechanisms remain relatively underdeveloped. As shown in
Figure~\ref{fig:Rec-Defense}, existing approaches can be broadly organized into proactive and post-hoc methods. Proactive methods integrate privacy protection into model development or training, including differential privacy\cite{mullner2023differential,yuan2023interaction}, regularization\cite{nasr2018machine,10.1145/3109859.3109906}, and popularity randomization\cite{zhang2021membership}. Post-hoc methods address privacy risks after model training, including machine unlearning
\cite{chen2022recommendation,li2023selective,NEURIPS2023_29a0ea49}, prompt-based defenses \cite{wen2024membershipinferenceattacksincontext,10.1145/3773078.3831733},
and agent-based defenses \cite{10.1145/3773078.3831733}. In addition, privacy risk estimation \cite{10.1145/3705328.3748052} provides complementary information for assessing membership leakage and guiding targeted protection decisions.

\subsection{Proactive Methods}
The proactive methods are dominated by differential privacy, while a few studies have also used regularization methods to address the model overfitting problem, which is believed to be the root cause of MIA.

\subsubsection{Differential Privacy}
Differential privacy (DP) \cite{abadi2016deep} is a rigorous probabilistic mechanism that provides information-theoretic guarantees of privacy: when a machine learning model is trained under a suitably small privacy budget, it cannot reliably learn or remember any single user’s data if the privacy budget is sufficiently small. The definition is that A (possibly randomized) mechanism $M$ is said to satisfy $(\varepsilon,\delta)$\nobreakdash-differential privacy (DP)\cite{abadi2016deep,kairouz2015composition,mullner2023differential} if, for all pairs of adjacent datasets $\mathcal{D}$ and $\mathcal{D}'$ differing in exactly one record, and for all measurable output events $S$, the following holds:
\begin{equation}
\Pr[M(\mathcal{D}) \in S] \;\le\; e^{\varepsilon}\, \Pr[M(\mathcal{D}') \in S] + \delta.
\label{eq:dp-def}
\end{equation}
This inequality ensures that the inclusion or exclusion of any single record has only a limited effect on the mechanism's output distribution, thereby providing a rigorous privacy guarantee. Furthermore, the MIA effectiveness level can be theoretically linked to and bounded by the privacy budget of DP, $\varepsilon$, as shown in later discussion, which is the unique strength of DP. 

In the context of RecSys, DP has been applied to mitigate membership inference attacks (MIAs) by adding calibrated noise to inputs, gradients, or model outputs~\cite{zhong2024interaction,10.1145/3726302.3730086}. Local differential privacy (LDP) is a client-side variant in which each user independently perturbs their data or model updates before transmitting them, thereby protecting against server-side inference. In federated RecSys, for instance, LDP has been evaluated as a defense strategy against interaction-level MIAs~\cite{yuan2023interaction}, showing that unless extremely large noise is applied the attacker's accuracy remains high, while excessive noise severely degrades recommendation performance. These findings highlight a fundamental trade-off: although DP and LDP offer formal membership-privacy guarantees, their practical deployment in RecSys is constrained by the privacy–utility dilemma. For example, while using LDP on a federated recommendation system reduces the effectiveness of attacks by up to 67\%, it also decreases the recommendation accuracy by 65\% \cite{yuan2023interaction}; on a centralized recommendation system\cite{zhong2024interaction}, when the defense effectiveness reaches 40\%, the model performance drops by 90\%. Designing defense mechanisms that preserve recommendation quality while providing tangible membership-privacy protection thus remains a critical open problem.

\subsubsection{Regularization}
Regularization aims to reduce the overfitting degree of target models to mitigate MIAs. Therefore, regularization methods that can reduce the overfitting of ML models can be leveraged to defend against MIAs. Existing regularization methods, including L2-norm regularization\cite{shokri17}, dropout \cite{srivastava2014dropout, salem2018mlleaksmodeldataindependent}, data augmentation\cite{pmlr-v139-kaya21a}, model stacking\cite{wu2022personalizednewsrecommendationmethods}, early stopping\cite{song2021systematic}, label smoothing \cite{szegedy2016rethinking}, adversarial regularization \cite{nasr2018machine}, and Mixup + MMD (Maximum Mean Discrepancy) \cite{li2021membership}, have been proposed and investigated as defense methods against MIAs in other fields \cite{song2021systematic,matsumoto2023,carlini2022lira,shokri17}. 

However, the study of using regularization as a defense method in RecSys is limited. Zhong et al. \cite{zhong2024interaction} firstly proposes regularization via gradient-level learning (RGL). The key idea is to introduce a regularization term into the target RecSys training objective to reduce the distinguishability between member and non-member samples from the perspective of the surrogate attacker. The regularization term is defined using the Kullback–Leibler (KL) divergence between the probability distributions of attack predictions for members and non-members. By penalizing the discrepancy between these distributions, RGL effectively mitigates the information gap exploited by MIAs and increase the defense efficiency. For example, RGL can increase the defense efficiency up to 37.8\% under MINER attack.

In the domain of social RecSys\cite{10.1145/3726302.3730086}, memorization of training data can inadvertently expose sensitive social relationships between users via recommendation outputs. Unlike many other application areas, social RecSys relies critically on user-to-user links to enhance recommendation quality. To mitigate this risk, the concept of Socially Adversarial Learning (SAL) has been developed specifically for the recommendation field. Under SAL, a surrogate attacker $\mathcal A'$ is embedded into model training: the attacker learns to distinguish user pairs $(u_1, u_2)$ that share a social link ($\mathcal U^+$) from those that do not ($\mathcal U^-$). The RecSys is then optimized with the combined objective
\[
\mathcal L_{\rm all} = \mathcal L_{\rm rec} + \lambda\,\mathcal L_{\rm def},
\]

where

\[
\mathcal L_{\rm def} = \mathrm{Dis}\bigl(\mathcal A'(\mathcal U^+),\,\mathcal A'(\mathcal U^-)\bigr)
\approx \sqrt{(\mu^+ - \mu^-)^2 + (\sigma^+ - \sigma^-)^2},
\]
in which $\mu^+, \mu^-$ and $\sigma^+, \sigma^-$ denote the means and standard deviations of the surrogate attacker’s output distributions for linked and non-linked user pairs, respectively. By minimizing this divergence term, SAL forces user-pair embeddings and recommendation outcomes for socially linked and unlinked users to become indistinguishable — thereby reducing the risk of social-relationship leakage.

Except for centralized RecSys, regularization has been used in Federated RecSys to protect user privacy information. To defend against interaction-level MIAs in Federated RecSys, Yuan et al. \cite{yuan2023interaction} proposed a regularization-style update control mechanism. Observing that user embeddings change little during training and that the public parameter updates from each client can leak membership signals, the approach augments the client-side optimization objective with a penalty term:
\[
\mathcal L = \mathcal L^{\rm rec} + \mu\,\|V^{t} - V^{0}\|,
\]
where \(\mathcal L^{\rm rec}\) denotes the standard recommendation loss, \(V^{0}\) is the client's initial public parameter vector, \(V^{t}\) is the uploaded parameter update at round \(t\), and \(\mu\) is the regularization strength. By constraining clients' updates from drifting far from the initial state, the defense reduces the distinguishability of members versus non-members that an attacker can exploit, while preserving recommendation utility more effectively than straightforward local differential privacy (LDP) noise addition.

While regularization-based defenses have received more attention in recommendation systems than differential privacy, they still inevitably involve a trade-off between privacy protection and recommendation utility. For instance, RGL improves defense effectiveness by up to 37.8\%, but this gain comes at the cost of a substantial utility drop of up to 12.7\%. Similarly, SAL enhances defense efficiency by up to 6.0\%, yet reduces model performance by as much as 9.7\%. These results highlight a fundamental limitation of regularization-based defenses: improving robustness against membership inference attacks often comes at the expense of degrading recommendation quality. Designing regularization techniques that effectively address the MIA threat, while minimizing utility loss, remains a critical open challenge.

\subsubsection{Popularity Randomization} 
To mitigate membership inference attacks in RecSys, Zhang et al.~\cite{zhang2021membership} propose a defense mechanism called Popularity Randomization. The core idea is that non-member users are typically recommended the most popular items, making their latent feature vectors unusually similar and therefore easily distinguished from members. To counter this weakness, the system expands the candidate pool of popular items for non-members and then randomly selects a subset for recommendation, thereby increasing randomness in non-members' outputs. Formally, when issuing recommendations to non-member users, instead of always selecting the top-\(k\) popular items, the method chooses a larger set of top-\(N\) popular items and randomly picks \(k\) items from within that set. This randomization breaks the deterministic mapping of non-members to the most popular items and reduces the similarity in feature vectors between non-members and members, thereby lowering the distinguishability exploited by the attack. 
However, this defense is vulnerable to attacks that assume non-members receive recommendations dominated by globally popular items. Modern RecSys can effectively address cold-start issues, meaning that even non-members often obtain reasonably personalized recommendations. As a result, the underlying assumption of this defense may not always hold in practice. Moreover, recent MIA studies have demonstrated that personalized recommendations themselves can leak sensitive user information\,\cite{zhu2023membership}. Therefore, the practical feasibility of this defense strategy remains uncertain.

\subsection{Post-hoc Methods}
Post-hoc methods aim to mitigate privacy risks without requiring privacy-preserving modifications during the original model training process. Unlike proactive approaches, these methods can be applied after training or during inference. Machine unlearning removes the influence of designated training data in response to privacy requirements, such as the ``right to be forgotten''. For LLM-based RecSys, prompt-based and agent-based defenses provide additional protection by controlling model behavior and monitoring recommendation outputs at inference time, without modifying model parameters
\cite{10.1145/3773078.3831733}. These approaches offer different privacy--utility
trade-offs, although their effectiveness against recommendation-specific MIAs remains an open challenge.In addition to that, the concept of \emph{privacy risk estimation} is introduced to allow both users and the model builder to learn the privacy risk of each contributed item, which may be used to decide data items to be removed. Due to the fully preserved utility, post-hoc methods might be more accepted in practice.  

\subsubsection{Machine Unlearning}
Unlearning has become a widely adopted post-hoc defense method, enabling model owners to meet users' privacy requests after deployment. Following the notion of unlearning principles \cite{nguyen2022survey}, we further categorize recommendation unlearning techniques into exact unlearning and approximate unlearning, depending on whether the method fully or partially removes the influence of a user's data from the trained model.

\textbf{Exact Unlearning.} Exact unlearning follows a strict and complete definition of machine unlearning, aiming to fully eliminate the influence of designated data samples at the algorithmic level. Inspired by the SISA method \cite{bourtoule2020machineunlearning}, most exact recommendation unlearning methods adopt the ensemble retraining framework. This framework partitions the original dataset into multiple subsets, trains a sub-model on each subset, and aggregates these sub-models to form the final predictor—similar to an ensemble learning pipeline. To guarantee algorithmic completeness, each sub-model is typically designed to be identical to the original model in terms of architecture, hyper-parameters, and training configurations. This design enables efficient unlearning: when a user submits an unlearning request, only the sub-model trained on the subset containing the target data needs to be retrained, avoiding full retraining of the entire dataset and thereby substantially improving efficiency. 

Building upon SISA, Chen et al.\cite{chen2022recommendation} propose RecEraser, which introduces two key modifications tailored to recommendation tasks.
First, RecEraser employs a balanced clustering module for dataset partitioning, grouping similar users or items into the same subset so as to preserve collaborative effects—unlike the random partitioning strategy used in SISA.
Second, RecEraser incorporates an attention-based aggregation network that learns adaptive weights for combining sub-models. Compared to uniform averaging or majority voting in SISA, this weighted aggregation significantly improves recommendation performance. Despite its advantages, the ensemble retraining framework faces a fundamental trade-off between unlearning efficiency and model utility in recommendation settings. Increasing the number of data partitions improves unlearning efficiency, but it also weakens collaborative signals, thereby degrading recommendation quality.  To preserve more utility, Li et al. \cite{NEURIPS2023_29a0ea49} propose UltraRE, a lightweight extension of RecEraser. UltraRE introduces a novel balanced clustering algorithm based on optimal transport theory, improving both clustering quality and computational efficiency. Furthermore, UltraRE simplifies the attention module by replacing it with a logistic regression model, further enhancing overall efficiency. 

Also, aiming to preserve more model utility, LASER adopts sequential training instead of parallel training during model aggregation \cite{li2022makingrecommendersystemsforget}. Sequential training processes sub-models one after another, which helps preserve cross-subset collaborative patterns that may be lost during parallel training. LASER additionally integrates curriculum learning to optimize the ordering of data subsets, further enhancing predictive performance. However, sequential training inevitably reduces unlearning efficiency. To address this, LASER incorporates early stopping and parameter manipulation strategies to shorten retraining time while maintaining unlearning completeness. 

Exact unlearning provides strong privacy guarantees by ensuring that all traces of targeted user data are completely removed from the training process, thereby eliminating their influence on the final model.

\textbf{Approximate Unlearning.}
Exact unlearning requires the presence of the entire training data, which is expensive to maintain and may not be available in a running system. Approximate unlearning aim to \emph{approximate} the effect of exact unlearning. These approaches operate either from a parametric perspective, directly manipulating model parameters, or from a functional perspective, fine-tuning the model so that its behavior resembles that of a model retrained without the forgotten data.

A major class of parametric approaches is reverse unlearning\cite{li2023selective,10.1145/3583780.3614811,10.1145/3701763}, where the influence of the target data is estimated and subtracted from the model parameters. Influence-function-based methods estimate this effect using gradient and Hessian information, providing a closed-form update that avoids additional training. However, in recommendation systems with high-dimensional user--item embeddings, these methods face significant computational overhead and may suffer from estimation inaccuracies. To mitigate this, recent work selectively computes influence only for target embeddings or prunes less important parameters to reduce the cost. From a functional perspective, active unlearning methods fine-tune the model toward an unlearned solution. Representative approaches include fine-tuning on retained data\cite{liu2022forgettingfastrecommendersystems} and using flipped losses for forget items\cite{10386603}. These methods are faster than exact retraining and can be applied to an already trained model, but lack theoretical guarantees because their performance depends heavily on optimization stability and the design of the fine-tuning objective.

Overall, approximate unlearning offers greater efficiency and is more practical for large-scale RecSys, but its reliability regarding privacy protection has been questioned \cite{gu2025auditingapproximatemachineunlearning}.

\subsubsection{Prompt-Based Defenses}
Prompt-based defenses aim to mitigate privacy leakage by modifying inference-time instructions or prompt configurations without retraining the underlying model.
In the context of ICL, Wen et al.\cite{wen2024membershipinferenceattacksincontext}
investigated instruction-level defenses that discourage LLMs from revealing information about private in-context demonstrations. Beyond instruction-based protection, Duan et al.~
\cite{duan2024privacyriskincontextlearning} proposed prompt ensembling to reduce membership
distinguishability by aggregating predictions from multiple prompted models. Similar ensemble-based defenses have also been evaluated in retrieval-based ICL settings\cite{kulkarni2026membershipinferenceattacksretrieval}.

In ICL-based RecSys, He et al.\cite{10.1145/3773078.3831733} evaluated prompt sanitization against membership inference attacks targeting user interactions embedded in recommendation prompts. Their findings show that although carefully designed instructions can mitigate certain attacks, they remain insufficient against recommendation-specific MIAs, particularly those exploiting item memorization and recommendation inertia. These findings highlight the limitations of existing prompt-based defenses and motivate the development of privacy-preserving prompting strategies tailored to personalized recommendation.

\subsubsection{Agent-Based Defenses}
Agent-based defenses introduce an additional monitoring layer to detect and mitigate privacy leakage during inference. Such approaches have been explored in other domains to enforce privacy policies without modifying the underlying model. For example, GuardAgent
\cite{3780338.3783063} employs an independent guard agent to verify whether target agents comply with predefined safety and privacy requirements. Similarly, PrivacyChecker
\cite{wang-etal-2025-privacy} monitors information flows in LLM-powered agents to
prevent inappropriate disclosure of sensitive information.

In ICL-based RecSys, He et al.\cite{10.1145/3773078.3831733} investigated an output-filtering defense agent that evaluates whether generated recommendations appear to reveal information from private in-context demonstrations. When potential leakage is detected, the agent requests the recommender to regenerate its output without modifying the original prompt or model parameters. However, the effectiveness of such defenses remains limited against recommendation-specific MIAs. Membership leakage can arise not only from explicit disclosure of private data, but also from implicit recommendation patterns that are difficult for output-filtering agents to identify. These findings highlight the need for privacy-aware defense agents capable of detecting membership signals embedded in personalized recommendations.

\subsubsection{Privacy Risk Estimation}
Unlearning provides a principled mechanism for ensuring user privacy by enabling the complete removal of a user's data from the training corpus. From the user's perspective, this capability aligns with the ``right to be forgotten,'' allowing individuals to request that their personal information be permanently deleted from the model's training set. From the company's perspective, however, a central challenge lies in systematically quantifying the privacy risk associated with retaining or removing specific user data. In this context, He et al.\cite{10.1145/3705328.3748052} introduced the \textbf{privacy risk score} that offers a formal measure of how much sensitive information a model may inadvertently reveal about a user's data, thereby providing a quantitative foundation for evaluating and enforcing privacy guarantees in machine learning systems. The privacy risk score is inspired by the differential privacy. In Eq.~\eqref{eq:dp-def}, under the hypothesis-testing interpretation of differential privacy, let $S$ denote the rejection region used by an arbitrary statistical test (or distinguisher) attempting to decide whether the mechanism’s output originated from $\mathcal{D}$ or $\mathcal{D}'$. If $\mathrm{TPR}$ and $\mathrm{FPR}$ represent the true and false positive rates of this test, respectively, then the privacy guarantee in Eq.~\eqref{eq:dp-def} imposes the following constraint on the achievable tradeoff between them:
\begin{equation}
\mathrm{TPR} \;\le\; e^{\varepsilon}\, \mathrm{FPR} + \delta,
\label{eq:roc-bound}
\end{equation}
and symmetrically with the roles of $\mathcal{D}$ and $\mathcal{D}'$ swapped.  

Equation~\eqref{eq:roc-bound} provides an intuitive ROC-curve interpretation of differential privacy: a smaller $(\varepsilon,\delta)$ pair uniformly bounds the distinguishing power of any potential adversary, thereby limiting the success probability of all membership inference attacks (MIAs) against the mechanism. Note that $\delta$ is often very small, and thus, can be safely removed from Eq. \ref{eq:roc-bound}. 

Interpreting Eq.~\eqref{eq:roc-bound} from a pointwise perspective naturally leads to the notion of an \emph{empirical indistinguishability level} for a specific record $z$, without the mechanism is applied. For a fixed trained model instance, a calibrated and sufficiently powerful membership inference attack (MIA) yields empirical values of $(\mathrm{TPR}, \mathrm{FPR})$ at a chosen operating point. Rearranging Eq.~\eqref{eq:roc-bound} then motivates the definition of a sample-specific privacy score \cite{10.1145/3705328.3748052}:
\[
\hat{\varepsilon}_z \;\approx\; \log\!\left(\frac{\mathrm{TPR}}{\max\{\mathrm{FPR},\,\epsilon_{\tiny\text{num}}\}}\right),
\]
where a small $\epsilon_{\tiny\text{num}}$ ensures numerical stability. Intuitively, $\hat{\varepsilon}_z$ quantifies the empirical privacy leakage of an individual record: higher values indicate that the record is easier to distinguish between membership and non-membership, and thus more privacy-sensitive for that model instance.  

In the context of RecSys, MIAs instantiate this principle at various granularities—\emph{user-level} and \emph{interaction-level}—by leveraging observable signals such as predicted scores, item ranks, or conversational traces. Consequently, such calibrated MIAs provide a practical, per-sample proxy for assessing privacy risk in RecSys, aligning empirical vulnerability measurement with the formal hypothesis-testing interpretation of differential privacy. The model owner can use this privacy risk score as a guiding instruction to remove the sensitive data from their training data to achieve the defensive purpose.

Overall, existing privacy protection mechanisms address membership leakage at different stages of the recommendation pipeline. However, their effectiveness remains constrained by privacy--utility trade-offs, incomplete removal guarantees, and limited robustness against emerging MIAs in LLM-based RecSys. Developing more effective and generalizable defenses therefore remains an important research direction.

\section{Evaluation}
\label{sec:Evaluation}
In this section, we systematically review the datasets and recommendation models commonly employed in MIA studies, summarize representative evaluation metrics for reproducible and comparable privacy evaluations.

\subsection{Datasets}
\label{subsec:EvaluationDatasets}

MIAs against RecSys are commonly evaluated using publicly available recommendation datasets containing user--item interactions, ratings, reviews, or social relationships. These datasets cover various application domains, including movies, music, e-commerce, gaming, and social networks. Their diverse characteristics enable researchers to investigate membership leakage under different recommendation architectures and interaction patterns.

Table~\ref{tab:datasets-MIAs} summarizes the statistics of selected datasets commonly considered in RecSys privacy research. Since different studies may apply distinct preprocessing procedures, the reported statistics should be interpreted with respect to their corresponding dataset versions.

\textbf{MovieLens}~\cite{harper2015movielens}.
MovieLens is one of the most widely adopted benchmark datasets in recommendation research. It contains user--movie ratings and is available in multiple versions with different scales. For example, MovieLens-100K and MovieLens-1M contain approximately
100,000 and one million rating interactions, respectively. These datasets are frequently used to evaluate membership inference against collaborative filtering and sequential recommendation models.

\textbf{Amazon}~\cite{hou2024bridging}.
The Amazon recommendation datasets contain user reviews, ratings, and product metadata collected from different product categories. Representative subsets include Digital Music, Beauty, Books, and Cell Phones. These datasets provide heterogeneous
user--item interactions and are commonly used to evaluate recommendation performance across different item domains.

\textbf{LastFM}.
LastFM contains user--artist interactions collected from the Last.fm music platform, including listening histories and related behavioral information. It is frequently used for music recommendation and user preference modeling. 

\textbf{Steam}~\cite{steam_200k}.
The Steam dataset contains user--game interaction records collected from a digital game distribution platform. The Steam-200K version includes records associated with game purchases and playtime. Its gaming-related interaction patterns provide a benchmark for evaluating recommendation models and membership inference attacks in
a domain different from conventional movie and product recommendation.

\textbf{Ta-Feng}~\cite{tafeng_grocery_dataset}.
Ta-Feng is a supermarket transaction dataset collected from a retail chain in Taiwan.
It contains transaction histories involving users, products, purchase quantities, and timestamps.The temporal information in these transactions makes the dataset suitable for investigating sequential recommendation and membership inference based on purchasing behavior.

\textbf{Yelp}~\cite{yelp_open_dataset}.
The Yelp dataset contains user reviews, ratings, and business metadata collected from a location-based review platform. It provides user--business interactions for evaluating recommendation models and privacy attacks. Different studies employ distinct releases or filtered subsets of Yelp, including the commonly used Yelp2018 recommendation benchmark. Consequently, dataset statistics may vary substantially depending on preprocessing and filtering criteria.

\textbf{Ciao}~\cite{tang_trust_web}.
The Ciao dataset contains user ratings, product information, and explicit social relationships, including trust links between users. The combination of user--item interactions and user--user relationships makes it particularly suitable for evaluating MIAs against social recommender systems, where membership leakage may concern both interaction records and social connections.

\textbf{Flickr}\footnote{\label{flickr_note}
\url{https://www.flickr.com/}}.
Flickr is a social image-sharing platform containing user interactions with images,
including favorites, comments, and tags, as well as user--user social connections.
Datasets derived from Flickr have been used to investigate socially aware recommendation and privacy risks associated with behavioral and social information.

\textbf{Book-Crossing and Delicious}.
Beyond conventional recommendation benchmarks, recent MIAs against LLM-based RecSys have considered additional datasets. Book-Crossing contains user--book rating interactions, whereas Delicious provides user interactions with online resources and tagging information.

\textbf{Douban}.
Douban is another representative dataset employed in cross-domain recommendation studies. It provides user preferences across different content domains and supports the evaluation of knowledge transfer between domains.

\begin{table*}[t]
\centering
\caption{Statistics of selected datasets used in
recommendation and membership inference studies.}
\label{tab:datasets-MIAs}

\footnotesize
\setlength{\tabcolsep}{6pt}
\renewcommand{\arraystretch}{1.18}
\sisetup{
    group-separator={,},
    group-minimum-digits=4
}

\begin{tabular*}{\textwidth}{
@{\extracolsep{\fill}}
l
S[table-format=8.0]
S[table-format=8.0]
S[table-format=9.0]
@{}
}
\toprule
\textbf{Dataset} &
{\textbf{Users}} &
{\textbf{Items}} &
{\textbf{Interactions}} \\
\midrule

MovieLens-100K~\cite{harper2015movielens}
& 943 & 1682 & 100000 \\

MovieLens-1M~\cite{harper2015movielens}
& 6040 & 3706 & 1000209 \\

Yelp~\cite{yelp_open_dataset}
& 1987897 & 150346 & 6990280 \\

Ta-Feng~\cite{tafeng_grocery_dataset}
& 32266 & 23812 & 817742 \\

Amazon-Digital Music~\cite{hou2024bridging}
& 100952 & 70519 & 130434 \\

Amazon-Beauty~\cite{hou2024bridging}
& 631986 & 115709 & 701528 \\

Amazon-Book~\cite{hou2024bridging}
& 10297355 & 4493336 & 29475453 \\

Amazon-Cell Phone~\cite{hou2024bridging}
& 11598197 & 1623399 & 20812945 \\

LastFM
& 23566 & 48123 & 1474122 \\

Steam~\cite{steam_200k}
& 12393 & 5155 & 200000 \\

Ciao~\cite{tang_trust_web}
& 7375 & 99746 & 278483 \\

Flickr
& 3074947 & 41278715 & 187168754 \\

\bottomrule
\end{tabular*}
\end{table*}

\subsection{Models}
\label{subsec:EvaluationModels}

The effectiveness of MIAs depends not only on
the attack methodology but also on the
underlying recommendation architecture.
Different RecSys models encode user preferences
through distinct mechanisms, including
collaborative filtering, sequential modeling,
graph propagation, cross-domain knowledge
transfer, and language-model-based recommendation.
These architectural differences may influence
the membership signals available to adversaries.

To provide a systematic overview, we organize
representative target models into the following
categories.

\textbf{Traditional and Neural Collaborative
Filtering Models.}

\textit{ICF}~\cite{sarwar2001item}
(Item-Based Collaborative Filtering) estimates
item similarities from historical user--item
interactions and recommends items similar
to those previously preferred by a user.

\textit{LFM}~\cite{koren2009matrix}
(Latent Factor Model) represents users and
items in a shared latent space, allowing
user preferences to be modeled through
interactions between latent representations.

\textit{NCF}~\cite{he2017neural}
(Neural Collaborative Filtering) extends
traditional collaborative filtering by
employing neural networks to capture
nonlinear user--item interaction patterns.

\textbf{Sequential Recommendation Models.}

\textit{GRU4Rec}~\cite{hidasi2015gru4rec}
employs gated recurrent units to model
session-based user interactions and predict
subsequent item preferences from sequential
behavior.

\textit{BERT4Rec}~\cite{sun2019bert4rec}
uses bidirectional Transformer representations
and masked-item prediction to capture
contextual dependencies in user interaction
sequences.

\textit{STAMP}~\cite{liu2018stamp}
(Short-Term Attention/Memory Priority Model)
captures both general session interests
and short-term preferences through
attention and memory mechanisms.

\textit{NARM}~\cite{li2017narm}
(Neural Attentive Recommendation Machine)
combines recurrent sequence encoding with
attention mechanisms to capture sequential
preferences and the primary intent of
a user session.

\textbf{Graph-Based and Knowledge-Graph
Recommendation Models.}

\textit{LightGCN}~\cite{he2020lightgcn}
simplifies graph convolution for collaborative
filtering by propagating user and item
embeddings over the interaction graph
without feature transformations and
nonlinear activation functions.

\textit{CKE}~\cite{zhang2016cke}
(Collaborative Knowledge Base Embedding)
integrates collaborative filtering with
knowledge-graph embeddings to incorporate
structural and semantic information into
recommendation.

\textit{KGAT}~\cite{wang2019kgat}
(Knowledge Graph Attention Network)
employs attention-based information
propagation over a collaborative
knowledge graph to capture high-order
user--item relationships.

\textbf{Social Recommendation Models.}

\textit{DiffNet++}~\cite{wu2020diffnetpp}
models social influence and user preference
propagation through graph-based diffusion
over social and user--item relationships.

\textit{DESIGN}~\cite{tao2022design}
incorporates denoising and self-supervised
learning to improve the quality of social
representations and mitigate noise in
social recommendation.

\textit{GDMSR}~\cite{quan2023gdmsr}
employs graph denoising and multi-relational
representation learning to capture useful
social and interaction signals while
reducing irrelevant relationships.

These models are particularly relevant
to privacy evaluations involving social
relationships, since recommendation outputs
may expose information about user--user
connections in addition to user--item
interactions.

\textbf{Cross-Domain Recommendation Models.}
\textit{EMCDR}~\cite{3172077.3172231}
Embedding and Mapping for Cross-Domain Recommendation learns domain-specific user and item embeddings and establishes mapping functions between their latent spaces. By transferring user representations from a source domain to a target domain, EMCDR enables recommendation for users with limited target-domain interactions.

\textit{CMF}~\cite{10.1145/1401890.1401969}
Collective Matrix Factorization jointly factorizes interaction matrices from multiple related domains while sharing latent representations of common entities. This shared-factor learning mechanism enables knowledge transfer across domains and improves
recommendation performance under sparse interaction settings.

\textit{CLFM}~\cite{10.1007/978-3-642-40991-2_11}
Cluster-Level Latent Factor Model employs joint nonnegative matrix tri-factorization to capture rating patterns at the user and item cluster levels. Unlike approaches that assume
fully shared latent representations, CLFM learns both common and domain-specific rating patterns, allowing knowledge transfer while preserving differences between domains.

\textbf{LLM-Based Recommendation Models.}
Recent recommendation systems increasingly leverage large language models (LLMs) to incorporate natural-language understanding and semantic knowledge into personalized recommendation. From a privacy perspective, two representative settings deserve particular attention: fine-tuned LLM-based RecSys\cite{10.1609/aaai.v40i27.39449} and ICL-based RecSys\cite{10.1145/3773078.3831733}.

\subsection{Evaluation Metrics}
\label{subsec:EvaluationMetrics}

Membership inference attacks are generally
formulated as binary classification tasks,
where an adversary attempts to distinguish
members from non-members based on observable
signals from a target recommender system.
Existing studies report several complementary
evaluation metrics, including attack success
rate (ASR), AUC, F1-score, true positive rate
(TPR), false positive rate (FPR), and
membership advantage.

\textbf{Attack Success Rate (ASR).}

ASR measures the proportion of successful
attack attempts among all evaluated cases:
\[
\mathrm{ASR}
=
\frac{N_{\mathrm{successful}}}
{N_{\mathrm{evaluated}}}.
\]

For binary membership inference,
when a successful attack is defined
as correctly predicting the membership
status of an evaluated sample,
ASR is equivalent to classification
accuracy. However, the success criterion
should be clearly specified when
attack methods involve task-specific
objectives or restricted target sets.

\textbf{AUC (ROC-AUC).}
\cite{zhang2021membership,wang2022debiasing,
zhu2023membership,10.1145/3726302.3730086}

The Area Under the Receiver Operating
Characteristic Curve (AUC) evaluates
the ability of an attack to distinguish
members from non-members across
different decision thresholds. It can be interpreted as the probability that a randomly selected member receives
a higher membership score than a randomly selected non-member, with ties receiving
appropriate treatment. An AUC of 0.5
indicates random discrimination,
whereas an AUC of 1 represents
perfect separation. AUC is widely used because it
summarizes discrimination performance
without requiring a fixed decision
threshold. However, it does not
fully characterize privacy risks
at operationally relevant
low false positive rates.

\textbf{F1-score.}
\cite{yuan2023interaction}

The F1-score is the harmonic mean
of precision and recall:
\[
\mathrm{F1}
=
\frac{
2 \cdot \mathrm{Precision}
\cdot \mathrm{Recall}
}{
\mathrm{Precision}
+
\mathrm{Recall}
}.
\]

F1-score balances precision
and recall at a selected
decision threshold.
It is particularly useful
when the relative importance
of false positives and
false negatives must be
considered simultaneously. However, F1-score depends
on the underlying class
distribution and should
therefore be interpreted
together with other metrics,
especially when the proportions
of members and non-members differ
between evaluation settings.

\textbf{TPR and FPR.}
\cite{He_2025,zhong2024interaction,carlini2022lira}

The True Positive Rate (TPR)
measures the proportion of
correctly identified members,
while the False Positive Rate
(FPR) measures the proportion
of non-members incorrectly
classified as members:
\[
\mathrm{TPR}
=
\frac{\mathrm{TP}}
{\mathrm{TP}+\mathrm{FN}},
\qquad
\mathrm{FPR}
=
\frac{\mathrm{FP}}
{\mathrm{FP}+\mathrm{TN}}.
\]

For membership inference,
achieving a high TPR at a
low FPR is particularly
important because it
indicates the ability
to identify members
while limiting false
membership accusations. Although AUC summarizes
overall attack discrimination,
it may obscure performance
in the low-FPR regime.
Following Carlini et al.~
\cite{carlini2022lira},
we recommend reporting TPR
at fixed FPR thresholds,
such as 0.1\% and 1\%,
alongside AUC, when the
evaluation size supports
reliable estimation.

\textbf{Membership Advantage.}
\cite{yeom2018privacy,10.1145/3773078.3831733}

Membership advantage measures
the difference between the
true positive rate and false
positive rate of an attack:
\[
\mathrm{Advantage}
=
\mathrm{TPR}
-
\mathrm{FPR}.
\]

Under a balanced evaluation
setting with equal numbers
of members and non-members,
this is equivalent to:
\[
\mathrm{Advantage}
=
2\left(
\mathrm{Accuracy}
-
0.5
\right).
\]

An advantage of 0 corresponds
to random-guessing performance,
while an advantage of 1 indicates
perfect membership inference.
A negative advantage indicates
that the attack performs worse
than random guessing under its
current decision rule. Higher positive advantage values
indicate stronger membership
distinguishability and greater
empirical leakage under the
evaluated attack setting.

Overall, no single metric
fully characterizes membership
privacy risks. We therefore
recommend reporting both
threshold-independent metrics,
such as AUC, and low-FPR
performance measures,
alongside clearly specified
decision thresholds and
class distributions.

\section{Discussion and Future Directions}
\label{sec:future and direction}

In this section, we discuss the main challenges and potential research opportunities in MIAs on RecSys.

\subsection{Adversary Modeling}

In practical scenarios, black-box RecSys models most closely reflect the adversary's perspective, where attackers can exploit only limited external information to compromise user privacy. The existence of such feasible black-box attacks significantly undermines user trust in service providers and poses long-term risks to corporate reputation and sustainability. Therefore, while the community continues to explore potential attack strategies and quantify the privacy risks associated with black-box models, equal emphasis should be placed on developing effective defense mechanisms. Given that even the most rigorous differential privacy (DP) approaches often lead to substantial utility degradation---rendering them less attractive for real-world deployment---future research should focus on designing adaptive and lightweight defense frameworks that balance privacy protection with model performance and business applicability.

In contrast, white-box settings assume that attackers have access to a model's internal parameters, gradients, or architecture, which is rarely achievable in practice. Nonetheless, white-box analyses provide valuable insights from the enterprise's perspective by quantifying internal privacy leakage and informing proactive protection strategies. The recently proposed notion of \emph{privacy risk scores}, which links membership inference attacks (MIAs) with differential privacy theory, offers a promising avenue for enterprises to identify and ``preemptively unlearn'' privacy-sensitive users or interactions before deployment. Developing more efficient and interpretable methods for estimating such risk scores---while ensuring minimal performance compromise---represents a compelling future direction toward achieving privacy-preserving yet high-utility RecSys.

\subsection{Importance of Interaction-Level MIAs}

Compared with user-level MIAs, interaction-level MIAs pose more significant privacy concerns, as they can reveal specific user behaviors and sensitive preferences rather than merely identifying a user's participation in model training. Therefore, understanding and mitigating interaction-level membership leakage is particularly important for protecting user privacy in RecSys. Recent work such as RecLiRA~\cite{10.1145/3705328.3748052} has explored quantifying membership leakage through interaction-level privacy risk scores, offering a promising foundation for targeted privacy protection. In particular, combining fine-grained privacy risk estimation with selective machine unlearning represents a valuable research direction, enabling RecSys providers to identify and remove high-risk interactions while minimizing the impact on recommendation utility.

However, the effectiveness of such risk-guided protection depends on whether the estimated privacy scores capture genuine membership leakage rather than ordinary preference generalization. This motivates further investigation into \emph{counterfactual interaction-level membership inference}. In RecSys, a target interaction may receive a high prediction score even when it was never included in training, simply because the user exhibits similar preferences or the item is popular among related users. Consequently, existing attacks may exploit correlations between recommendation behavior and membership rather than signals attributable to the inclusion of a specific interaction. Future research could investigate counterfactual evaluation frameworks that compare recommendation behavior when a target interaction is included or excluded from training, while controlling for user preferences, item popularity, and collaborative dependencies. Such evaluations would help isolate membership-specific signals and improve the reliability of interaction-level privacy risk assessment.

Beyond preference generalization, the construction of interaction data introduces another challenge: \emph{exposure-aware membership inference}. In implicit-feedback RecSys, unobserved interactions do not necessarily indicate negative user preferences, as an item may never have been exposed to the user. Moreover, negative sampling introduces additional complexity because sampled negative items may participate in model optimization without representing actual user interactions. These distinctions raise an important question of whether existing MIAs can reliably distinguish positive interaction membership from exposure patterns, negative sampling artifacts, and general preference signals. Future studies should investigate membership inference under different interaction observation and sampling mechanisms, using evaluation protocols that distinguish training-data inclusion from interaction existence. Such research could clarify how exposure bias and negative sampling strategies influence membership leakage across recommendation architectures.

Furthermore, the challenges of interaction-level membership inference extend beyond individual user--item pairs, particularly in sequential RecSys. Existing membership definitions often treat interactions as isolated records, overlooking their temporal dependencies and behavioral contexts. For example, the same user may interact with an item multiple times across different sessions, making pair-level membership insufficient to distinguish individual behavioral events. This motivates \emph{sequence-dependent membership inference}, which investigates whether specific interaction events or subsequences were included in model training. Future research could examine how interaction ordering, temporal dependencies, and session boundaries influence membership distinguishability, thereby extending interaction-level privacy analysis from static user--item relationships to dynamic behavioral patterns.

Together, these directions highlight the need to move beyond improving attack accuracy toward understanding the underlying sources and granularity of interaction-level membership leakage. Developing counterfactual, exposure-aware, and sequence-sensitive evaluation frameworks could provide a more reliable foundation for fine-grained privacy risk estimation and targeted protection in RecSys.

\subsection{Relationship Between User-Level and Interaction-Level MIAs}

Understanding the relationship between user-level and interaction-level membership leakage is essential for developing comprehensive privacy protection mechanisms in RecSys. Recent work, such as RecPS~\cite{10.1145/3705328.3748052}, has extended interaction-level privacy scores to the user level and demonstrated the effectiveness of risk-guided interaction removal. Moreover, its experiments revealed the \emph{privacy onion effect}\cite{carlini21Onion}, where removing high-risk users may increase the privacy scores of some remaining users. These findings suggest that membership privacy risks are interdependent rather than isolated across users and interactions.

An important direction is to investigate how correlated interactions jointly contribute to user-level membership leakage. Although aggregating interaction-level privacy scores provides a practical approximation, it may not fully capture the dependencies among interactions or the distinguishability of an entire user's training history. In particular, combinations of interactions may produce membership signals that differ from those suggested by their individual privacy scores. Future research could investigate dependency-aware privacy composition methods that account for collaborative relationships, interaction correlations, and heterogeneous user histories, providing more reliable estimates of user-level membership risks.

Furthermore, the privacy onion effect raises a fundamental question: \emph{why does protecting certain users sometimes increase the privacy risks of others?} Since RecSys learn collaborative representations from interconnected user--item interactions, removing a subset of training records may alter the influence of the remaining records on the learned model. Future studies could investigate how user similarity, shared item embeddings, interaction sparsity, and graph connectivity contribute to such cross-user privacy effects. Understanding these mechanisms may enable the prediction of which users are most likely to experience increased membership leakage following data removal.

These findings could motivate a broader direction toward \emph{system-wide privacy-aware unlearning}. Rather than optimizing privacy protection exclusively for users requesting data removal, future methods could consider the resulting changes in membership risks across the remaining user population. Developing unlearning strategies that jointly minimize targeted privacy leakage, limit adverse privacy effects on other users, and preserve recommendation utility would provide a more comprehensive foundation for privacy protection in collaborative RecSys.

\subsection{New Attack Targets: Social-Level MIAs}

Unlike user-level and interaction-level privacy threats, the risk posed by users' social relationships introduces a novel vantage point on the serious privacy dangers inherent in modern RecSys, especially on social media. Current work largely limits itself to detecting whether two users share a social connection (e.g., mutual following, friendship, or ``likes''). In real-world settings, however, a more meaningful dimension of social privacy lies in the private nature of those connections---for example, relationships with close friends or family members. A public celebrity follow may appear innocuous, but the inference that a user is connected to a close friend or family member potentially exposes far greater personal risk. Thus, a promising future direction is the formulation of more granular social-relation membership inference attacks (MIAs) and the development of corresponding defenses that protect fine-grained user-pair social privacy without degrading recommendation utility.

\subsection{Attacks on Emerging RecSys Models}

Despite recent advances in MIAs against RecSys, membership privacy risks in federated, cross-domain, and multimodal recommendation architectures remain insufficiently explored. Existing studies have primarily established the feasibility of membership inference under specific models and adversarial assumptions, leaving many architecture-specific leakage mechanisms and corresponding protection strategies underexplored. These emerging RecSys paradigms introduce distinct privacy challenges through decentralized optimization, cross-domain knowledge transfer, and multimodal information fusion.

In \emph{Federated RecSys}, existing MIAs exploit shared model updates to infer private user--item interactions, as discussed in Section~\ref{sec:fed-mias}. However, negative sampling introduces an important challenge: both positive interactions and sampled negative items may participate in local optimization, making their contributions difficult to distinguish from shared updates. Future research could investigate \emph{negative-sampling-aware membership inference}, particularly under adaptive or unknown sampling strategies, to better distinguish positive interaction membership from negative-sample participation. Such investigations could further motivate privacy-preserving sampling and update mechanisms that conceal private interactions while preserving recommendation utility. Moreover, graph-based Federated RecSys introduce additional opportunities to investigate membership leakage associated with private social relations through shared graph representations.

In \emph{Cross-Domain RecSys}, existing research~\cite{11036029} has demonstrated that transferring user preferences across domains can expose membership information. However, the mechanisms through which membership signals propagate between domains remain insufficiently understood. A promising direction is to identify whether membership leakage originates from shared user representations, cross-domain alignment, or transferred interaction patterns, particularly when attackers can access only target-domain recommendation outputs. Understanding these leakage pathways could motivate \emph{membership-aware knowledge transfer}, where privacy-sensitive information is selectively protected during representation alignment or knowledge transfer. An important challenge is determining whether source-domain membership exposure can be reduced without sacrificing the benefits of cross-domain personalization, especially in data-sparse and cold-start settings.

Meanwhile, \emph{Multimodal RecSys} introduce additional membership privacy challenges by integrating heterogeneous data sources, including text, images, and audio. Although multimodal representations enhance recommendation quality, their fusion may also alter the distinguishability of private user interactions. Future research could investigate whether information fusion amplifies membership leakage compared with individual modalities, and whether sensitive membership information associated with one modality can be inferred through another. Understanding these cross-modal dependencies could provide a foundation for privacy-aware multimodal representation learning and fusion mechanisms that mitigate membership leakage while retaining complementary recommendation signals.

\subsection{Membership Privacy Across the LLM-Based Recommendation}
\label{sec:llm_membership_future}

Unlike conventional RecSys, LLM-based recommendation pipelines may incorporate private user information through multiple components, including tokenizers, model training, persistent memory, retrieval, and inference-time prompts. Existing studies have investigated membership inference against in-context recommendation~\cite{he2025membershipinferenceattacksllmbased} and training-data membership using knowledge distillation~\cite{10.1609/aaai.v40i27.39449}. However, the diverse mechanisms through which private information is processed and retained introduce additional membership privacy challenges. Understanding how membership signals originate, propagate, and persist across these components represents an important research direction.

\textbf{Provenance-Aware Membership Inference.}
A fundamental challenge is determining what constitutes membership when the same user interaction may contribute to different stages of personalization. Unlike conventional training-data membership, private interactions may also be incorporated into persistent user memories or dynamically selected inference contexts. These membership sources are conceptually distinct but may produce overlapping recommendation signals. A promising direction is to develop a provenance-aware formulation that distinguishes membership according to the stage at which private information is incorporated. For a target interaction $z$ at time $t$, a possible formulation is
\[
\mathbf{m}(z,t)=
\left(
m_{\mathrm{train}}(z),
m_{\mathrm{memory}}(z,t),
m_{\mathrm{context}}(z,t)
\right),
\]
where the components indicate whether the target interaction contributed to model training, persistent memory, or the current inference context, respectively. Such contributions may include derived information, such as summarized user preferences, rather than exact copies of the original interaction. Future research could investigate whether these membership sources can be distinguished through observable recommendation behavior, and whether membership signals originating from different components reinforce or obscure one another. This would extend conventional binary membership inference toward a more comprehensive characterization of privacy leakage across personalization stages.

\textbf{Tokenizer-Level Membership Leakage.}
Beyond model parameters and inference-time contexts, tokenizers represent another potential source of membership leakage in LLM-based RecSys. Tong et al.~\cite{317733} demonstrated membership inference against LLM tokenizers, while Hayase et al.~\cite{NEURIPS2024_10e6dfea} showed that BPE tokenizers can reveal information about training corpus composition. Although the latter concerns data mixture inference rather than instance-level membership, these findings motivate investigating whether tokenizers trained or adapted on recommendation-specific corpora encode membership signals associated with private user interactions. An important distinction is whether a recommender adopts a fixed pretrained tokenizer or adapts its tokenizer using recommendation data, since a tokenizer that remains unchanged cannot directly memorize subsequently introduced interactions. Future research could investigate the conditions under which tokenizer-level membership leakage arises and whether privacy-preserving tokenizer adaptation can mitigate such risks without compromising recommendation performance.

\textbf{Query-Dependent Membership Exposure.}
In retrieval-augmented recommendation, the presence of private interactions in persistent storage does not necessarily imply their inclusion in a particular inference context. Instead, retrieval decisions depend on the current recommendation query, user preferences, and candidate items. Inspired by recent membership inference research on retrieval-augmented generation~\cite{zeng-etal-2025-mitigating}, future studies could investigate whether carefully constructed recommendation queries expose membership information associated with specific retrieval events. In particular, an important research question is whether attackers can distinguish the presence of an interaction in the underlying retrieval corpus from its actual inclusion in a personalized prompt. Understanding this distinction could motivate retrieval mechanisms that limit membership-dependent information exposure while preserving the benefits of personalized recommendations.

\textbf{Persistent and Cross-Stage Membership Leakage.}
Another challenge arises when private user interactions are transformed into persistent profiles, summarized memories, or intermediate representations. Removing an interaction from its original storage location may not eliminate its influence if derived information remains accessible to subsequent personalization stages. Furthermore, information derived from one user's interactions may influence other users' recommendations when shared retrieval resources or aggregated representations are employed. Future research could investigate how membership signals persist or propagate through memory updates, retrieval operations, and prompt construction, and whether removing private information from one component sufficiently prevents leakage through others. Understanding these dependencies could motivate privacy-preserving memory management and cross-stage unlearning mechanisms that address the propagation of sensitive information throughout the recommendation pipeline.

\textbf{Pipeline-Level Privacy Evaluation and Protection.}
These challenges highlight the need for evaluation frameworks that consider membership leakage across the entire LLM-based recommendation lifecycle. Existing evaluation settings commonly focus on individual attack surfaces, making it difficult to identify the origin of observed membership signals. Future benchmarks could independently control whether target interactions contribute to model training, persistent memory, retrieval, and inference-time context construction. Such controlled evaluations could help distinguish component-specific leakage from leakage arising through interactions among multiple components, while accounting for ordinary preference generalization. Beyond attack effectiveness, future research should investigate how privacy interventions at different stages affect downstream membership leakage and recommendation utility. Establishing provenance-aware evaluation protocols would provide a foundation for developing end-to-end privacy protection mechanisms tailored to LLM-based RecSys.

\subsection{Trade-offs in Defense Methods}

Existing defenses against MIAs in RecSys, including differential privacy (DP), regularization, and machine unlearning, face an inherent trade-off between privacy protection and recommendation utility. As discussed in Section~\ref{sec:Defense}, DP provides formal privacy guarantees under specified assumptions but may degrade recommendation quality, whereas regularization and approximate unlearning often offer more flexible protection without equivalent guarantees. Rather than applying uniform privacy mechanisms to all training data, future research should investigate more targeted and adaptive defense strategies that account for heterogeneous membership risks and recommendation-specific data dependencies.

One promising direction is \emph{privacy-spillover-aware protection}. Although RecPS~\cite{10.1145/3705328.3748052} has demonstrated the potential of risk-guided unlearning, the privacy onion effect indicates that removing high-risk data may increase membership leakage for some remaining users. This motivates defense mechanisms that consider privacy consequences beyond the targeted interactions. For example, future approaches could combine privacy risk estimation with iterative unlearning and post-removal auditing to identify unintended increases in other users' privacy risks. An important challenge is to minimize such privacy spillovers while preserving recommendation quality and respecting required data removal requests.

Another direction is \emph{architecture-specific privacy protection}. As discussed in the preceding subsections, different RecSys architectures expose membership information through distinct mechanisms. In Federated RecSys, privacy-aware update aggregation or perturbation strategies could be designed to reduce the distinguishability of positive interactions from sampled negatives. In Cross-Domain RecSys, membership-aware representation alignment or selective knowledge transfer could suppress privacy-sensitive signals while retaining transferable user preferences. Similarly, Multimodal RecSys could benefit from modality-specific privacy constraints that limit leakage through shared representations. These approaches motivate investigating whether targeted protection of membership-sensitive components can achieve better privacy--utility trade-offs than applying uniform defenses to the entire model.

Finally, emerging LLM-based RecSys call for \emph{lifecycle-wide privacy enforcement}. Since private interactions may contribute to model parameters, persistent memories, retrieval databases, and inference contexts, protecting a single component may leave membership signals accessible through others. A promising direction is to combine data provenance tracking, privacy-aware retrieval, and cross-component unlearning to manage sensitive information throughout the personalization lifecycle. Such mechanisms could enforce interaction-level protection policies across different stages and verify whether privacy interventions effectively reduce downstream membership leakage. Developing these coordinated defenses while minimizing computational overhead and preserving personalization quality remains an important research challenge.

\section{Conclusion}
\label{sec:conclusion}

Membership inference in RecSys is not a direct extension of conventional MIAs. Sparse interaction histories, shared user--item representations, ranked outputs, and heterogeneous training pipelines jointly determine whether and how private participation can be distinguished. By reviewing user-, interaction-, and social-relation-level attacks, this survey highlights why membership must be explicitly defined and separated from ordinary preference prediction or link inference. It also shows that observed attack effectiveness depends on both adversarial access and recommendation-specific data dependencies.

Our synthesis of defenses underscores that reducing empirical attack success is not equivalent to obtaining formal privacy guarantees, and that privacy interventions can alter risks for other users. These observations motivate research on counterfactual and exposure-aware membership evaluation, cross-user privacy dependencies, and architecture-specific safeguards for federated, cross-domain, and multimodal recommendation. LLM-based RecSys further require provenance-aware analysis across training, tokenization, persistent memory, retrieval, and inference contexts. Progress will depend on clearly defined membership targets, realistic threat models, and reproducible evaluations that jointly consider low-FPR attack effectiveness, recommendation utility, and defense costs. By connecting attack mechanisms, privacy assessment, and protection strategies, this survey aims to support more reliable and targeted membership privacy protection in evolving RecSys.

\bibliographystyle{ACM-Reference-Format}
\bibliography{papers}

@article{movielens,
	address = {New York, NY, USA},
	articleno = {19},
	author = {Harper, F. Maxwell and Konstan, Joseph A.},
	doi = {10.1145/2827872},
	issn = {2160-6455},
	issue_date = {January 2016},
	journal = {ACM Trans. Interact. Intell. Syst.},
	month = dec,
	number = {4},
	numpages = {19},
	publisher = {Association for Computing Machinery},
	title = {The MovieLens Datasets: History and Context},
	url = {https://doi.org/10.1145/2827872},
	volume = {5},
	year = {2015}}

@inproceedings{matsumoto2023,
	author = {Matsumoto, Tomoya and Miura, Takayuki and Yanai, Naoto},
	booktitle = {2023 IEEE Security and Privacy Workshops (SPW)},
	organization = {IEEE},
	pages = {77--83},
	title = {Membership inference attacks against diffusion models},
	year = {2023}}

@inproceedings{carlini21Onion,
	author = {Carlini, Nicholas and Jagielski, Matthew and Zhang, Chiyuan and Papernot, Nicolas and Terzis, Andreas and Tramer, Florian},
	booktitle = {Advances in Neural Information Processing Systems},
	editor = {S. Koyejo and S. Mohamed and A. Agarwal and D. Belgrave and K. Cho and A. Oh},
	pages = {13263--13276},
	publisher = {Curran Associates, Inc.},
	title = {The Privacy Onion Effect: Memorization is Relative},
	url = {https://proceedings.neurips.cc/paper_files/paper/2022/file/564b5f8289ba846ebc498417e834c253-Paper-Conference.pdf},
	volume = {35},
	year = {2022}}

@inproceedings{carlini2022lira,
	author = {Carlini, Nicholas and Chien, Steve and Nasr, Milad and Song, Shuang and Terzis, Andreas and Tramer, Florian},
	booktitle = {2022 IEEE Symposium on Security and Privacy (SP)},
	organization = {IEEE},
	pages = {1897--1914},
	title = {Membership inference attacks from first principles},
	year = {2022}}

@inproceedings{abadi2016deep,
	author = {Abadi, Martin and Chu, Andy and Goodfellow, Ian and McMahan, H Brendan and Mironov, Ilya and Talwar, Kunal and Zhang, Li},
	booktitle = {Proceedings of the 2016 ACM SIGSAC conference on computer and communications security},
	pages = {308--318},
	title = {Deep learning with differential privacy},
	year = {2016}}

@article{hu2022,
	author = {Hu, Hongsheng and Salcic, Zoran and Sun, Lichao and Dobbie, Gillian and Yu, Philip S and Zhang, Xuyun},
	journal = {ACM Computing Surveys (CSUR)},
	number = {11s},
	pages = {1--37},
	publisher = {ACM New York, NY},
	title = {Membership inference attacks on machine learning: A survey},
	volume = {54},
	year = {2022}}

@inproceedings{zhang2020,
	author = {Zhang, Yuheng and Jia, Ruoxi and Pei, Hengzhi and Wang, Wenxiao and Li, Bo and Song, Dawn},
	booktitle = {Proceedings of the IEEE/CVF conference on computer vision and pattern recognition},
	pages = {253--261},
	title = {The secret revealer: Generative model-inversion attacks against deep neural networks},
	year = {2020}}

@inproceedings{Choquette-Choo21,
	author = {Choquette-Choo, Christopher A and Tramer, Florian and Carlini, Nicholas and Papernot, Nicolas},
	booktitle = {International conference on machine learning},
	organization = {PMLR},
	pages = {1964--1974},
	title = {Label-only membership inference attacks},
	year = {2021}}

@inproceedings{Yeom18,
	author = {Yeom, Samuel and Giacomelli, Irene and Fredrikson, Matt and Jha, Somesh},
	booktitle = {2018 IEEE 31st computer security foundations symposium (CSF)},
	organization = {IEEE},
	pages = {268--282},
	title = {Privacy risk in machine learning: Analyzing the connection to overfitting},
	year = {2018}}

@inproceedings{shokri17,
	author = {Shokri, Reza and Stronati, Marco and Song, Congzheng and Shmatikov, Vitaly},
	booktitle = {2017 IEEE symposium on security and privacy (SP)},
	organization = {IEEE},
	pages = {3--18},
	title = {Membership inference attacks against machine learning models},
	year = {2017}}

@inproceedings{zhu2023membership,
	author = {Zhu, Zhihao and Wu, Chenwang and Fan, Rui and Lian, Defu and Chen, Enhong},
	booktitle = {Proceedings of the ACM Web Conference 2023},
	pages = {1208--1219},
	title = {Membership inference attacks against sequential recommender systems},
	year = {2023}}

@article{li2023selective,
	author = {Li, Yuyuan and Chen, Chaochao and Zheng, Xiaolin and Zhang, Yizhao and Gong, Biao and Wang, Jun and Chen, Linxun},
	journal = {Expert Systems with Applications},
	pages = {121025},
	publisher = {Elsevier},
	title = {Selective and collaborative influence function for efficient recommendation unlearning},
	volume = {234},
	year = {2023}}

@inproceedings{yuan2023interaction,
	author = {Yuan, Wei and Yang, Chaoqun and Nguyen, Quoc Viet Hung and Cui, Lizhen and He, Tieke and Yin, Hongzhi},
	booktitle = {Proceedings of the ACM Web Conference 2023},
	pages = {1053--1062},
	title = {Interaction-level membership inference attack against federated recommender systems},
	year = {2023}}

@article{mullner2023differential,
	author = {M{\"u}llner, Peter and Lex, Elisabeth and Schedl, Markus and Kowald, Dominik},
	journal = {Frontiers in big Data},
	pages = {1249997},
	publisher = {Frontiers Media SA},
	title = {Differential privacy in collaborative filtering recommender systems: a review},
	volume = {6},
	year = {2023}}

@inproceedings{chen2022recommendation,
	author = {Chen, Chong and Sun, Fei and Zhang, Min and Ding, Bolin},
	booktitle = {Proceedings of the ACM Web Conference 2022},
	pages = {2768--2777},
	title = {Recommendation unlearning},
	year = {2022}}

@inproceedings{zhang2021membership,
	author = {Zhang, Minxing and Ren, Zhaochun and Wang, Zihan and Ren, Pengjie and Chen, Zhunmin and Hu, Pengfei and Zhang, Yang},
	booktitle = {Proceedings of the 2021 ACM SIGSAC Conference on Computer and Communications Security},
	pages = {864--879},
	title = {Membership inference attacks against recommender systems},
	year = {2021}}

@inproceedings{wang2022debiasing,
	author = {Wang, Zihan and Huang, Na and Sun, Fei and Ren, Pengjie and Chen, Zhumin and Luo, Hengliang and de Rijke, Maarten and Ren, Zhaochun},
	booktitle = {Proceedings of the 28th ACM SIGKDD Conference on Knowledge Discovery and Data Mining},
	pages = {1959--1968},
	title = {Debiasing learning for membership inference attacks against recommender systems},
	year = {2022}}

@inproceedings{sun2019bert4rec,
	author = {Sun, Fei and Liu, Jun and Wu, Jian and Pei, Changhua and Lin, Xiao and Ou, Wenwu and Jiang, Peng},
	booktitle = {Proceedings of the 28th ACM international conference on information and knowledge management},
	pages = {1441--1450},
	title = {BERT4Rec: Sequential recommendation with bidirectional encoder representations from transformer},
	year = {2019}}

@inproceedings{he2020lightgcn,
	author = {He, Xiangnan and Deng, Kuan and Wang, Xiang and Li, Yan and Zhang, Yongdong and Wang, Meng},
	booktitle = {Proceedings of the 43rd International ACM SIGIR conference on research and development in Information Retrieval},
	pages = {639--648},
	title = {Lightgcn: Simplifying and powering graph convolution network for recommendation},
	year = {2020}}

@inproceedings{he2017neural,
	author = {He, Xiangnan and Liao, Lizi and Zhang, Hanwang and Nie, Liqiang and Hu, Xia and Chua, Tat-Seng},
	booktitle = {Proceedings of the 26th international conference on world wide web},
	pages = {173--182},
	title = {Neural collaborative filtering},
	year = {2017}}

@inproceedings{kairouz2015composition,
	author = {Kairouz, Peter and Oh, Sewoong and Viswanath, Pramod},
	booktitle = {International conference on machine learning},
	organization = {PMLR},
	pages = {1376--1385},
	title = {The composition theorem for differential privacy},
	year = {2015}}

@inproceedings{song2021systematic,
  title={Systematic evaluation of privacy risks of machine learning models},
  author={Song, Liwei and Mittal, Prateek},
  booktitle={30th USENIX Security Symposium (USENIX Security 21)},
  pages={2615--2632},
  year={2021}
}

@article{nguyen2022survey,
  title={A survey of machine unlearning},
  author={Nguyen, Thanh Tam and Huynh, Thanh Trung and Ren, Zhao and Nguyen, Phi Le and Liew, Alan Wee-Chung and Yin, Hongzhi and Nguyen, Quoc Viet Hung},
  journal={arXiv preprint arXiv:2209.02299},
  year={2022}
}

@inproceedings{sablayrolles2019white,
  title={White-box vs black-box: Bayes optimal strategies for membership inference},
  author={Sablayrolles, Alexandre and Douze, Matthijs and Schmid, Cordelia and Ollivier, Yann and J{\'e}gou, Herv{\'e}},
  booktitle={International Conference on Machine Learning},
  pages={5558--5567},
  year={2019},
  organization={PMLR}
}

@inproceedings{shokri2017membership,
  title={Membership inference attacks against machine learning models},
  author={Shokri, Reza and Stronati, Marco and Song, Congzheng and Shmatikov, Vitaly},
  booktitle={2017 IEEE symposium on security and privacy (SP)},
  pages={3--18},
  year={2017},
  organization={IEEE}
}

@inproceedings{yeom2018privacy,
  title={Privacy risk in machine learning: Analyzing the connection to overfitting},
  author={Yeom, Samuel and Giacomelli, Irene and Fredrikson, Matt and Jha, Somesh},
  booktitle={2018 IEEE 31st computer security foundations symposium (CSF)},
  pages={268--282},
  year={2018},
  organization={IEEE}
}

@inproceedings{zhong2024interaction,
  title={Interaction-level Membership Inference Attack against Recommender Systems with Long-tailed Distribution},
  author={Zhong, Da and Wang, Xiuling and Xu, Zhichao and Xu, Jun and Wang, Wendy Hui},
  booktitle={Proceedings of the 33rd ACM International Conference on Information and Knowledge Management},
  pages={3433--3442},
  year={2024}
}

@inproceedings{Nasr_2019,
   title={Comprehensive Privacy Analysis of Deep Learning: Passive and Active White-box Inference Attacks against Centralized and Federated Learning},
   url={http://dx.doi.org/10.1109/SP.2019.00065},
   DOI={10.1109/sp.2019.00065},
   booktitle={2019 IEEE Symposium on Security and Privacy (SP)},
   publisher={IEEE},
   author={Nasr, Milad and Shokri, Reza and Houmansadr, Amir},
   year={2019},
   month=may, pages={739–753} }

@misc{liu2022membershipinferenceattacksexploiting,
      title={Membership Inference Attacks by Exploiting Loss Trajectory}, 
      author={Yiyong Liu and Zhengyu Zhao and Michael Backes and Yang Zhang},
      year={2022},
      eprint={2208.14933},
      archivePrefix={arXiv},
      primaryClass={cs.CR},
      url={https://arxiv.org/abs/2208.14933}, 
}

@misc{wen2024membershipinferenceattacksincontext,
      title={Membership Inference Attacks Against In-Context Learning}, 
      author={Rui Wen and Zheng Li and Michael Backes and Yang Zhang},
      year={2024},
      eprint={2409.01380},
      archivePrefix={arXiv},
      primaryClass={cs.CR},
      url={https://arxiv.org/abs/2409.01380}, 
}

@article{Zhao_2024,
   title={Recommender Systems in the Era of Large Language Models (LLMs)},
   volume={36},
   ISSN={2326-3865},
   url={http://dx.doi.org/10.1109/TKDE.2024.3392335},
   DOI={10.1109/tkde.2024.3392335},
   number={11},
   journal={IEEE Transactions on Knowledge and Data Engineering},
   publisher={Institute of Electrical and Electronics Engineers (IEEE)},
   author={Zhao, Zihuai and Fan, Wenqi and Li, Jiatong and Liu, Yunqing and Mei, Xiaowei and Wang, Yiqi and Wen, Zhen and Wang, Fei and Zhao, Xiangyu and Tang, Jiliang and Li, Qing},
   year={2024},
   month=nov, pages={6889–6907} }

@misc{gao2023chatrecinteractiveexplainablellmsaugmented,
      title={Chat-REC: Towards Interactive and Explainable LLMs-Augmented Recommender System}, 
      author={Yunfan Gao and Tao Sheng and Youlin Xiang and Yun Xiong and Haofen Wang and Jiawei Zhang},
      year={2023},
      eprint={2303.14524},
      archivePrefix={arXiv},
      primaryClass={cs.IR},
      url={https://arxiv.org/abs/2303.14524}, 
}

@misc{chen2023knowledgegraphcompletionmodels,
      title={Knowledge Graph Completion Models are Few-shot Learners: An Empirical Study of Relation Labeling in E-commerce with LLMs}, 
      author={Jiao Chen and Luyi Ma and Xiaohan Li and Nikhil Thakurdesai and Jianpeng Xu and Jason H. D. Cho and Kaushiki Nag and Evren Korpeoglu and Sushant Kumar and Kannan Achan},
      year={2023},
      eprint={2305.09858},
      archivePrefix={arXiv},
      primaryClass={cs.IR},
      url={https://arxiv.org/abs/2305.09858}, 
}

@inproceedings{10.1145/3543507.3583355,
author = {Chen, Xiao and Fan, Wenqi and Chen, Jingfan and Liu, Haochen and Liu, Zitao and Zhang, Zhaoxiang and Li, Qing},
title = {Fairly Adaptive Negative Sampling for Recommendations},
year = {2023},
isbn = {9781450394161},
publisher = {Association for Computing Machinery},
address = {New York, NY, USA},
url = {https://doi.org/10.1145/3543507.3583355},
doi = {10.1145/3543507.3583355},
booktitle = {Proceedings of the ACM Web Conference 2023},
pages = {3723–3733},
numpages = {11},
location = {Austin, TX, USA},
series = {WWW '23}
}

@inproceedings{Bao_2023, series={RecSys ’23},
   title={TALLRec: An Effective and Efficient Tuning Framework to Align Large Language Model with Recommendation},
   url={http://dx.doi.org/10.1145/3604915.3608857},
   DOI={10.1145/3604915.3608857},
   booktitle={Proceedings of the 17th ACM Conference on Recommender Systems},
   publisher={ACM},
   author={Bao, Keqin and Zhang, Jizhi and Zhang, Yang and Wang, Wenjie and Feng, Fuli and He, Xiangnan},
   year={2023},
   month=sep, pages={1007–1014},
   collection={RecSys ’23} }

@misc{cui2022m6recgenerativepretrainedlanguage,
      title={M6-Rec: Generative Pretrained Language Models are Open-Ended Recommender Systems}, 
      author={Zeyu Cui and Jianxin Ma and Chang Zhou and Jingren Zhou and Hongxia Yang},
      year={2022},
      eprint={2205.08084},
      archivePrefix={arXiv},
      primaryClass={cs.IR},
      url={https://arxiv.org/abs/2205.08084}, 
}

@inproceedings{10.1145/3523227.3546767,
author = {Geng, Shijie and Liu, Shuchang and Fu, Zuohui and Ge, Yingqiang and Zhang, Yongfeng},
title = {Recommendation as Language Processing (RLP): A Unified Pretrain, Personalized Prompt \& Predict Paradigm (P5)},
year = {2022},
isbn = {9781450392785},
publisher = {Association for Computing Machinery},
address = {New York, NY, USA},
url = {https://doi.org/10.1145/3523227.3546767},
doi = {10.1145/3523227.3546767},
booktitle = {Proceedings of the 16th ACM Conference on Recommender Systems},
pages = {299–315},
numpages = {17},
location = {Seattle, WA, USA},
series = {RecSys '22}
}

@inproceedings{Wang_2019, series={KDD ’19},
   title={KGAT: Knowledge Graph Attention Network for Recommendation},
   url={http://dx.doi.org/10.1145/3292500.3330989},
   DOI={10.1145/3292500.3330989},
   booktitle={Proceedings of the 25th ACM SIGKDD International Conference on Knowledge Discovery \& Data Mining},
   publisher={ACM},
   author={Wang, Xiang and He, Xiangnan and Cao, Yixin and Liu, Meng and Chua, Tat-Seng},
   year={2019},
   month=jul, pages={950–958},
   collection={KDD ’19} }

@misc{duan2024privacyriskincontextlearning,
      title={On the Privacy Risk of In-context Learning}, 
      author={Haonan Duan and Adam Dziedzic and Mohammad Yaghini and Nicolas Papernot and Franziska Boenisch},
      year={2024},
      eprint={2411.10512},
      archivePrefix={arXiv},
      primaryClass={cs.LG},
      url={https://arxiv.org/abs/2411.10512}, 
}

@article{hou2024bridging,
  title={Bridging Language and Items for Retrieval and Recommendation},
  author={Hou, Yupeng and Li, Jiacheng and He, Zhankui and Yan, An and Chen, Xiusi and McAuley, Julian},
  journal={arXiv preprint arXiv:2403.03952},
  year={2024}
}

@inproceedings{10.1145/3705328.3748052,
author = {He, Jiajie and Gu, Yuechun and Chen, Keke},
title = {RecPS: Privacy Risk Scoring for Recommender Systems},
year = {2025},
isbn = {9798400713644},
publisher = {Association for Computing Machinery},
address = {New York, NY, USA},
url = {https://doi.org/10.1145/3705328.3748052},
doi = {10.1145/3705328.3748052},
booktitle = {Proceedings of the Nineteenth ACM Conference on Recommender Systems},
pages = {432–440},
numpages = {9},
location = {
},
series = {RecSys '25}
}

@misc{chi2024shadowfreemembershipinferenceattacks,
      title={Shadow-Free Membership Inference Attacks: Recommender Systems Are More Vulnerable Than You Thought}, 
      author={Xiaoxiao Chi and Xuyun Zhang and Yan Wang and Lianyong Qi and Amin Beheshti and Xiaolong Xu and Kim-Kwang Raymond Choo and Shuo Wang and Hongsheng Hu},
      year={2024},
      eprint={2405.07018},
      archivePrefix={arXiv},
      primaryClass={cs.CR},
      url={https://arxiv.org/abs/2405.07018}, 
}

@inproceedings{shah21_interspeech,
  title     = {Evaluating the Vulnerability of End-to-End Automatic Speech Recognition Models to Membership Inference Attacks},
  author    = {Muhammad A. Shah and Joseph Szurley and Markus Mueller and Athanasios Mouchtaris and Jasha Droppo},
  year      = {2021},
  booktitle = {Interspeech 2021},
  pages     = {891--895},
  doi       = {10.21437/Interspeech.2021-1188},
  issn      = {2958-1796},
}

@inproceedings{He_2025, series={RecSys ’25},
   title={RecPS: Privacy Risk Scoring for Recommender Systems},
   url={http://dx.doi.org/10.1145/3705328.3748052},
   DOI={10.1145/3705328.3748052},
   booktitle={Proceedings of the Nineteenth ACM Conference on Recommender Systems},
   publisher={ACM},
   author={He, Jiajie and Gu, Yuechun and Chen, Keke},
   year={2025},
   month=sep, pages={432–440},
   collection={RecSys ’25} }

@misc{wu2022personalizednewsrecommendationmethods,
      title={Personalized News Recommendation: Methods and Challenges}, 
      author={Chuhan Wu and Fangzhao Wu and Yongfeng Huang and Xing Xie},
      year={2022},
      eprint={2106.08934},
      archivePrefix={arXiv},
      primaryClass={cs.IR},
      url={https://arxiv.org/abs/2106.08934}, 
}

@inproceedings{10.1145/3726302.3730086,
author = {Zhao, Xuhao and Zhang, Zhongrui and Zhu, Yanmin and Wang, Zhaobo and Ma, Wenze and Yu, Jiadi and Tang, Feilong},
title = {Social Relation-Level Privacy Risks and Preservation in Social Recommender Systems},
year = {2025},
isbn = {9798400715921},
publisher = {Association for Computing Machinery},
address = {New York, NY, USA},
url = {https://doi.org/10.1145/3726302.3730086},
doi = {10.1145/3726302.3730086},
booktitle = {Proceedings of the 48th International ACM SIGIR Conference on Research and Development in Information Retrieval},
pages = {1728–1737},
numpages = {10},
location = {Padua, Italy},
series = {SIGIR '25}
}

@misc{rendle2012bprbayesianpersonalizedranking,
      title={BPR: Bayesian Personalized Ranking from Implicit Feedback}, 
      author={Steffen Rendle and Christoph Freudenthaler and Zeno Gantner and Lars Schmidt-Thieme},
      year={2012},
      eprint={1205.2618},
      archivePrefix={arXiv},
      primaryClass={cs.IR},
      url={https://arxiv.org/abs/1205.2618}, 
}

@misc{mao2023simplexsimplestrongbaseline,
      title={SimpleX: A Simple and Strong Baseline for Collaborative Filtering}, 
      author={Kelong Mao and Jieming Zhu and Jinpeng Wang and Quanyu Dai and Zhenhua Dong and Xi Xiao and Xiuqiang He},
      year={2023},
      eprint={2109.12613},
      archivePrefix={arXiv},
      primaryClass={cs.IR},
      url={https://arxiv.org/abs/2109.12613}, 
}

@inproceedings{10.1145/3658644.3690335,
author = {Li, Hao and Li, Zheng and Wu, Siyuan and Hu, Chengrui and Ye, Yutong and Zhang, Min and Feng, Dengguo and Zhang, Yang},
title = {SeqMIA: Sequential-Metric Based Membership Inference Attack},
year = {2024},
isbn = {9798400706363},
publisher = {Association for Computing Machinery},
address = {New York, NY, USA},
url = {https://doi.org/10.1145/3658644.3690335},
doi = {10.1145/3658644.3690335},
booktitle = {Proceedings of the 2024 on ACM SIGSAC Conference on Computer and Communications Security},
pages = {3496–3510},
numpages = {15},
location = {Salt Lake City, UT, USA},
series = {CCS '24}
}

@article{harper2015movielens,
  title={The movielens datasets: History and context},
  author={Harper, F Maxwell and Konstan, Joseph A},
  journal={Acm transactions on interactive intelligent systems (tiis)},
  volume={5},
  number={4},
  pages={1--19},
  year={2015},
  publisher={Acm New York, NY, USA}
}

@misc{yelp_open_dataset,
  author = {Yelp Inc.},
  title = {Yelp Open Dataset},
  year = {2025},
  howpublished = {\url{https://business.yelp.com/data/resources/open-dataset/}},
  note = {Accessed: 2025-11-06}
}

@misc{tafeng_grocery_dataset,
  author = {Chiranjiv Das},
  title = {Ta Feng Grocery Dataset},
  year = {2018},
  howpublished = {\url{https://www.kaggle.com/datasets/chiranjivdas09/ta-feng-grocery-dataset}},
  note = {Accessed: 2025-11-06}
}

@misc{steam_200k,
    title = {Steam Video Games Dataset},
    author = {Tamber},
    year = {2017},
    howpublished = {\url{https://www.kaggle.com/datasets/tamber/steam-video-games}},
    note = {Accessed: 2025-11-06}
}

@misc{ammaduddin2019federatedcollaborativefilteringprivacypreserving,
      title={Federated Collaborative Filtering for Privacy-Preserving Personalized Recommendation System}, 
      author={Muhammad Ammad-ud-din and Elena Ivannikova and Suleiman A. Khan and Were Oyomno and Qiang Fu and Kuan Eeik Tan and Adrian Flanagan},
      year={2019},
      eprint={1901.09888},
      archivePrefix={arXiv},
      primaryClass={cs.IR},
      url={https://arxiv.org/abs/1901.09888}, 
}

@inproceedings{10.1145/3394486.3403176,
author = {Muhammad, Khalil and Wang, Qinqin and O'Reilly-Morgan, Diarmuid and Tragos, Elias and Smyth, Barry and Hurley, Neil and Geraci, James and Lawlor, Aonghus},
title = {FedFast: Going Beyond Average for Faster Training of Federated Recommender Systems},
year = {2020},
isbn = {9781450379984},
publisher = {Association for Computing Machinery},
address = {New York, NY, USA},
url = {https://doi.org/10.1145/3394486.3403176},
doi = {10.1145/3394486.3403176},
booktitle = {Proceedings of the 26th ACM SIGKDD International Conference on Knowledge Discovery \& Data Mining},
pages = {1234–1242},
numpages = {9},
location = {Virtual Event, CA, USA},
series = {KDD '20}
}

@misc{tang_trust_web,
  author = {Jiliang Tang},
  title = {Trust/Distrust Computing},
  howpublished = {\url{https://www.cse.msu.edu/~tangjili/trust.html}},
  institution = {Michigan State University},
  year = {2014},
  note = {Accessed: 2025-11-07}
}

@inproceedings{hidasi2015gru4rec,
  title={Session-based Recommendations with Recurrent Neural Networks},
  author={Hidasi, Bal{\'a}zs and Karatzoglou, Alexandros and Baltrunas, Linas and Tikk, Domonkos},
  booktitle={International Conference on Learning Representations (ICLR)},
  year={2016}
}

@inproceedings{liu2018stamp,
  title={STAMP: Short-Term Attention/Memory Priority Model for Session-based Recommendation},
  author={Liu, Qiao and Zeng, Yifu and Mokhosi, Ria and Zhang, Haibin},
  booktitle={Proceedings of the 24th ACM SIGKDD International Conference on Knowledge Discovery \& Data Mining},
  pages={1831--1839},
  year={2018}
}

@inproceedings{li2017narm,
  title={Neural Attentive Session-based Recommendation},
  author={Li, Jing and Ren, Pengjie and Chen, Zhumin and Ren, Zhaochun and Lian, Defu and Ma, Shengxian and de Rijke, Maarten},
  booktitle={Proceedings of the 2017 ACM on Conference on Information and Knowledge Management (CIKM)},
  pages={1449--1458},
  year={2017},
  organization={ACM}
}

@inproceedings{zhang2016cke,
  title={Collaborative Knowledge Base Embedding for Recommender Systems},
  author={Zhang, Fuzheng and Yuan, Nicholas Jing and Lian, Defu and Xie, Xing and Ma, Wei-Ying},
  booktitle={Proceedings of the 22nd ACM SIGKDD International Conference on Knowledge Discovery and Data Mining},
  pages={353--362},
  year={2016},
  organization={ACM}
}

@inproceedings{wang2019kgat,
  title={KGAT: Knowledge Graph Attention Network for Recommendation},
  author={Wang, Xiang and He, Xiangnan and Cao, Yixin and Liu, Meng and Chua, Tat-Seng},
  booktitle={Proceedings of the 25th ACM SIGKDD International Conference on Knowledge Discovery \& Data Mining},
  pages={950--958},
  year={2019},
  organization={ACM}
}

@article{wu2020diffnetpp,
  title={DiffNet++: A Neural Influence and Interest Diffusion Network for Social Recommendation},
  author={Wu, Le and Li, Junwei and Sun, Peijie and Hong, Richang and Ge, Yong and Wang, Meng},
  journal={IEEE Transactions on Knowledge and Data Engineering},
  year={2020}
}

@inproceedings{quan2023gdmsr,
  title={Robust Preference-Guided Denoising for Graph based Social Recommendation},
  author={Quan, Yuhan and Ding, Jingtao and Gao, Chen and Yi, Lingling and Jin, Depeng and Li, Yong},
  booktitle={Proceedings of the ACM Web Conference 2023 (WWW)},
  year={2023}
}

@inproceedings{sarwar2001item,
  title={Item-based collaborative filtering recommendation algorithms},
  author={Sarwar, Badrul and Karypis, George and Konstan, Joseph and Riedl, John},
  booktitle={Proceedings of the 10th international conference on World Wide Web},
  pages={285--295},
  year={2001}
}

@inproceedings{koren2009matrix,
  title={Matrix Factorization Techniques for Recommender Systems},
  author={Koren, Yehuda and Bell, Robert and Volinsky, Chris},
  booktitle={Computer},
  volume={42},
  number={8},
  pages={30--37},
  year={2009},
  organization={IEEE}
}

@inproceedings{tao2022design,
  title={Revisiting graph based social recommendation: A distillation enhanced social graph network},
  author={Tao, Ye and Li, Ying and Zhang, Su and Hou, Zhirong and Wu, Zhonghai},
  booktitle={Proceedings of the ACM Web Conference 2022},
  pages={2830--2838},
  year={2022}
}

@inproceedings{melis2019exploiting,
  title={Exploiting unintended feature leakage in collaborative learning},
  author={Melis, Luca and Song, Congzheng and De Cristofaro, Emiliano and Shmatikov, Vitaly},
  booktitle={2019 IEEE symposium on security and privacy (SP)},
  pages={691--706},
  year={2019},
  organization={IEEE}
}

@InProceedings{pmlr-v54-mcmahan17a,
  title = 	 {{Communication-Efficient Learning of Deep Networks from Decentralized Data}},
  author = 	 {McMahan, Brendan and Moore, Eider and Ramage, Daniel and Hampson, Seth and Arcas, Blaise Aguera y},
  booktitle = 	 {Proceedings of the 20th International Conference on Artificial Intelligence and Statistics},
  pages = 	 {1273--1282},
  year = 	 {2017},
  editor = 	 {Singh, Aarti and Zhu, Jerry},
  volume = 	 {54},
  series = 	 {Proceedings of Machine Learning Research},
  month = 	 {20--22 Apr},
  publisher =    {PMLR},
  url = 	 {https://proceedings.mlr.press/v54/mcmahan17a.html}
}

@misc{mcmahan2018learningdifferentiallyprivaterecurrent,
      title={Learning Differentially Private Recurrent Language Models}, 
      author={H. Brendan McMahan and Daniel Ramage and Kunal Talwar and Li Zhang},
      year={2018},
      eprint={1710.06963},
      archivePrefix={arXiv},
      primaryClass={cs.LG},
      url={https://arxiv.org/abs/1710.06963}, 
}

@article{10.1145/3560486,
author = {Imran, Mubashir and Yin, Hongzhi and Chen, Tong and Nguyen, Quoc Viet Hung and Zhou, Alexander and Zheng, Kai},
title = {ReFRS: Resource-efficient Federated Recommender System for Dynamic and Diversified User Preferences},
year = {2023},
issue_date = {July 2023},
publisher = {Association for Computing Machinery},
address = {New York, NY, USA},
volume = {41},
number = {3},
issn = {1046-8188},
url = {https://doi.org/10.1145/3560486},
doi = {10.1145/3560486},
journal = {ACM Trans. Inf. Syst.},
month = feb,
articleno = {65},
numpages = {30}
}

@misc{wang2021fastadaptingprivacypreservingfederatedrecommender,
      title={Fast-adapting and Privacy-preserving Federated Recommender System}, 
      author={Qinyong Wang and Hongzhi Yin and Tong Chen and Junliang Yu and Alexander Zhou and Xiangliang Zhang},
      year={2021},
      eprint={2104.00919},
      archivePrefix={arXiv},
      primaryClass={cs.IR},
      url={https://arxiv.org/abs/2104.00919}, 
}

@misc{zhang2023comprehensiveprivacyanalysisfederated,
      title={Comprehensive Privacy Analysis on Federated Recommender System against Attribute Inference Attacks}, 
      author={Shijie Zhang and Wei Yuan and Hongzhi Yin},
      year={2023},
      eprint={2205.11857},
      archivePrefix={arXiv},
      primaryClass={cs.IR},
      url={https://arxiv.org/abs/2205.11857}, 
}

@ARTICLE{9170754,
  author={Lin, Guanyu and Liang, Feng and Pan, Weike and Ming, Zhong},
  journal={IEEE Intelligent Systems}, 
  title={FedRec: Federated Recommendation With Explicit Feedback}, 
  year={2021},
  volume={36},
  number={5},
  pages={21-30},
  doi={10.1109/MIS.2020.3017205}}

@inproceedings{10.1145/3460231.3478855,
author = {Lin, Zhaohao and Pan, Weike and Ming, Zhong},
title = {FR-FMSS: Federated Recommendation via Fake Marks and Secret Sharing},
year = {2021},
isbn = {9781450384582},
publisher = {Association for Computing Machinery},
address = {New York, NY, USA},
url = {https://doi.org/10.1145/3460231.3478855},
doi = {10.1145/3460231.3478855},
booktitle = {Proceedings of the 15th ACM Conference on Recommender Systems},
pages = {668–673},
numpages = {6},
location = {Amsterdam, Netherlands},
series = {RecSys '21}
}

@article{10.1145/3548456,
author = {Lin, Zhaohao and Pan, Weike and Yang, Qiang and Ming, Zhong},
title = {A Generic Federated Recommendation Framework via Fake Marks and Secret Sharing},
year = {2022},
issue_date = {April 2023},
publisher = {Association for Computing Machinery},
address = {New York, NY, USA},
volume = {41},
number = {2},
issn = {1046-8188},
url = {https://doi.org/10.1145/3548456},
doi = {10.1145/3548456},
journal = {ACM Trans. Inf. Syst.},
month = dec,
articleno = {40},
numpages = {37}
}

@INPROCEEDINGS {8594844,
author = { Kang, Wang-Cheng and McAuley, Julian },
booktitle = { 2018 IEEE International Conference on Data Mining (ICDM) },
title = {{ Self-Attentive Sequential Recommendation }},
year = {2018},
volume = {},
ISSN = {},
pages = {197-206},
doi = {10.1109/ICDM.2018.00035},
url = {https://doi.ieeecomputersociety.org/10.1109/ICDM.2018.00035},
publisher = {IEEE Computer Society},
address = {Los Alamitos, CA, USA},
month =Nov}

@inproceedings{tang2018personalized,
  author = {Tang, Jiaxi and Wang, Ke},
  booktitle = {Proceedings of the Eleventh {ACM} International Conference on Web Search and Data Mining - {WSDM} {\textquotesingle}18},
  doi = {10.1145/3159652.3159656},
  publisher = {{ACM} Press},
  title = {Personalized Top-N Sequential Recommendation via Convolutional Sequence Embedding},
  year = 2018
}

@article{srivastava2014dropout,
  title={Dropout: a simple way to prevent neural networks from overfitting},
  author={Srivastava, Nitish and Hinton, Geoffrey and Krizhevsky, Alex and Sutskever, Ilya and Salakhutdinov, Ruslan},
  journal={The journal of machine learning research},
  volume={15},
  number={1},
  pages={1929--1958},
  year={2014},
  publisher={JMLR. org}
}

@inproceedings{szegedy2016rethinking,
  title={Rethinking the inception architecture for computer vision},
  author={Szegedy, Christian and Vanhoucke, Vincent and Ioffe, Sergey and Shlens, Jon and Wojna, Zbigniew},
  booktitle={Proceedings of the IEEE conference on computer vision and pattern recognition},
  pages={2818--2826},
  year={2016}
}

@inproceedings{nasr2018machine,
  title={Machine learning with membership privacy using adversarial regularization},
  author={Nasr, Milad and Shokri, Reza and Houmansadr, Amir},
  booktitle={Proceedings of the 2018 ACM SIGSAC conference on computer and communications security},
  pages={634--646},
  year={2018}
}

@inproceedings{li2021membership,
  title={Membership inference attacks and defenses in classification models},
  author={Li, Jiacheng and Li, Ninghui and Ribeiro, Bruno},
  booktitle={Proceedings of the Eleventh ACM Conference on Data and Application Security and Privacy},
  pages={5--16},
  year={2021}
}

@misc{bourtoule2020machineunlearning,
      title={Machine Unlearning}, 
      author={Lucas Bourtoule and Varun Chandrasekaran and Christopher A. Choquette-Choo and Hengrui Jia and Adelin Travers and Baiwu Zhang and David Lie and Nicolas Papernot},
      year={2020},
      eprint={1912.03817},
      archivePrefix={arXiv},
      primaryClass={cs.CR},
      url={https://arxiv.org/abs/1912.03817}, 
}

@inproceedings{NEURIPS2023_29a0ea49,
 author = {Li, Yuyuan and Chen, Chaochao and Zhang, Yizhao and Liu, Weiming and Lyu, Lingjuan and Zheng, Xiaolin and Meng, Dan and Wang, Jun},
 booktitle = {Advances in Neural Information Processing Systems},
 editor = {A. Oh and T. Naumann and A. Globerson and K. Saenko and M. Hardt and S. Levine},
 pages = {12611--12625},
 publisher = {Curran Associates, Inc.},
 title = {UltraRE: Enhancing RecEraser for Recommendation Unlearning via Error Decomposition},
 url = {https://proceedings.neurips.cc/paper_files/paper/2023/file/29a0ea49a103a233b17c0705cdeccb66-Paper-Conference.pdf},
 volume = {36},
 year = {2023}
}

@misc{li2022makingrecommendersystemsforget,
      title={Making Recommender Systems Forget: Learning and Unlearning for Erasable Recommendation}, 
      author={Yuyuan Li and Xiaolin Zheng and Chaochao Chen and Junlin Liu},
      year={2022},
      eprint={2203.11491},
      archivePrefix={arXiv},
      primaryClass={cs.IR},
      url={https://arxiv.org/abs/2203.11491}, 
}

@inproceedings{10.1145/3583780.3614811,
author = {Zhang, Shuijing and Lou, Jian and Xiong, Li and Zhang, Xiaoyu and Liu, Jing},
title = {Closed-form Machine Unlearning for Matrix Factorization},
year = {2023},
isbn = {9798400701245},
publisher = {Association for Computing Machinery},
address = {New York, NY, USA},
url = {https://doi.org/10.1145/3583780.3614811},
doi = {10.1145/3583780.3614811},
booktitle = {Proceedings of the 32nd ACM International Conference on Information and Knowledge Management},
pages = {3278–3287},
numpages = {10},
location = {Birmingham, United Kingdom},
series = {CIKM '23}
}

@article{10.1145/3701763,
author = {Zhang, Yang and Hu, Zhiyu and Bai, Yimeng and Wu, Jiancan and Wang, Qifan and Feng, Fuli},
title = {Recommendation Unlearning via Influence Function},
year = {2024},
issue_date = {June 2025},
publisher = {Association for Computing Machinery},
address = {New York, NY, USA},
volume = {3},
number = {2},
url = {https://doi.org/10.1145/3701763},
doi = {10.1145/3701763},
journal = {ACM Trans. Recomm. Syst.},
month = dec,
articleno = {22},
numpages = {23}
}

@misc{liu2022forgettingfastrecommendersystems,
      title={Forgetting Fast in Recommender Systems}, 
      author={Wenyan Liu and Juncheng Wan and Xiaoling Wang and Weinan Zhang and Dell Zhang and Hang Li},
      year={2022},
      eprint={2208.06875},
      archivePrefix={arXiv},
      primaryClass={cs.IR},
      url={https://arxiv.org/abs/2208.06875}, 
}

@INPROCEEDINGS{10386603,
  author={Alshehri, Manal A. and Zhang, Xiangliang},
  booktitle={2023 IEEE International Conference on Big Data (BigData)}, 
  title={Forgetting User Preference in Recommendation Systems with Label-Flipping}, 
  year={2023},
  volume={},
  number={},
  pages={271-281},
  doi={10.1109/BigData59044.2023.10386603}}

@misc{gu2025auditingapproximatemachineunlearning,
      title={Auditing Approximate Machine Unlearning for Differentially Private Models}, 
      author={Yuechun Gu and Jiajie He and Keke Chen},
      year={2025},
      eprint={2508.18671},
      archivePrefix={arXiv},
      primaryClass={cs.LG},
      url={https://arxiv.org/abs/2508.18671}, 
}

@inproceedings {317733,
author = {Meng Tong and Yuntao Du and Kejiang Chen and Weiming Zhang and Ninghui Li},
title = {Membership Inference Attacks on Tokenizers of Large Language Models},
booktitle = {35th USENIX Security Symposium (USENIX Security 26)},
year = {2026},
isbn = {978-1-939133-58-8},
address = {Baltimore, MD},
pages = {1207--1226},
url = {https://www.usenix.org/conference/usenixsecurity26/presentation/tong},
publisher = {USENIX Association},
month = aug
}

@inproceedings{NEURIPS2024_10e6dfea,
 author = {Hayase, Jonathan and Liu, Alisa and Choi, Yejin and Oh, Sewoong and Smith, Noah},
 booktitle = {Advances in Neural Information Processing Systems},
 doi = {10.52202/079017-0285},
 editor = {A. Globerson and L. Mackey and D. Belgrave and A. Fan and U. Paquet and J. Tomczak and C. Zhang},
 pages = {8956--8983},
 publisher = {Curran Associates, Inc.},
 title = {Data Mixture Inference Attack: BPE Tokenizers Reveal Training Data Compositions},
 url = {https://proceedings.neurips.cc/paper_files/paper/2024/file/10e6dfea9a673bef4a7b1cb9234891bc-Paper-Conference.pdf},
 volume = {37},
 year = {2024}
}

@INPROCEEDINGS{11036029,
  author={Yang, Jingru and Liu, Baisong},
  booktitle={2025 IEEE 6th International Seminar on Artificial Intelligence, Networking and Information Technology (AINIT)}, 
  title={Membership Inference Attacks Against Cross-Domain Recommender Systems}, 
  year={2025},
  volume={},
  number={},
  pages={932-938},
  doi={10.1109/AINIT65432.2025.11036029}}

@ARTICLE{11372209,
  author={Chen, Jinyin and Bai, Yang and Ma, Minying and Chen, Defu and Zheng, Yayu and Chen, Tieming and Ji, Baiyang},
  journal={IEEE Transactions on Computational Social Systems}, 
  title={Causal Inference-Based Collaborative Membership Inference Attack on Federated Recommendation}, 
  year={2026},
  volume={13},
  number={3},
  pages={3745-3762},
  doi={10.1109/TCSS.2026.3654406}}

@inproceedings{10.1609/aaai.v40i27.39449,
author = {Li, Cuihong and Huang, Xiaowen and Yin, Chuanhuan and Sang, Jitao},
title = {Membership inference attack against large language model-based recommendation systems: a new distillation-based paradigm},
year = {2026},
isbn = {978-1-57735-906-7},
publisher = {AAAI Press},
url = {https://doi.org/10.1609/aaai.v40i27.39449},
doi = {10.1609/aaai.v40i27.39449},
booktitle = {Proceedings of the Fortieth AAAI Conference on Artificial Intelligence and Thirty-Eighth Conference on Innovative Applications of Artificial Intelligence and Sixteenth Symposium on Educational Advances in Artificial Intelligence},
articleno = {2548},
numpages = {10},
series = {AAAI'26/IAAI'26/EAAI'26}
}

@inproceedings{10.1145/3404835.3463032,
author = {Wang, Jinpeng and Zhu, Jieming and He, Xiuqiang},
title = {Cross-Batch Negative Sampling for Training Two-Tower Recommenders},
year = {2021},
isbn = {9781450380379},
publisher = {Association for Computing Machinery},
address = {New York, NY, USA},
url = {https://doi.org/10.1145/3404835.3463032},
doi = {10.1145/3404835.3463032},
booktitle = {Proceedings of the 44th International ACM SIGIR Conference on Research and Development in Information Retrieval},
pages = {1632–1636},
numpages = {5},
location = {Virtual Event, Canada},
series = {SIGIR '21}
}

@inproceedings{10.1145/3773078.3831733,
author = {He, Jiajie and Chen, Min-Chun and Chen, Xintong and Fang, Xinyang and Gu, Yuechun and Chen, Keke},
title = {Membership Inference Attacks on In-Context Learning Recommendation},
year = {2026},
isbn = {9798400722844},
publisher = {Association for Computing Machinery},
address = {New York, NY, USA},
url = {https://doi.org/10.1145/3773078.3831733},
doi = {10.1145/3773078.3831733},
booktitle = {Proceedings of the 20th ACM Conference on Recommender Systems},
pages = {370–379},
numpages = {10},
location = {Minneapolis, USA},
series = {RecSys '26}
}

@article{10.1145/3568954,
author = {Xin, Xin and Yang, Jiyuan and Wang, Hanbing and Ma, Jun and Ren, Pengjie and Luo, Hengliang and Shi, Xinlei and Chen, Zhumin and Ren, Zhaochun},
title = {On the User Behavior Leakage from Recommender System Exposure},
year = {2023},
issue_date = {July 2023},
publisher = {Association for Computing Machinery},
address = {New York, NY, USA},
volume = {41},
number = {3},
issn = {1046-8188},
url = {https://doi.org/10.1145/3568954},
doi = {10.1145/3568954},
journal = {ACM Trans. Inf. Syst.},
month = feb,
articleno = {57},
numpages = {25}
}

@inproceedings{NEURIPS2019_60a6c400,
 author = {Zhu, Ligeng and Liu, Zhijian and Han, Song},
 booktitle = {Advances in Neural Information Processing Systems},
 editor = {H. Wallach and H. Larochelle and A. Beygelzimer and F. d\textquotesingle Alch\'{e}-Buc and E. Fox and R. Garnett},
 pages = {},
 publisher = {Curran Associates, Inc.},
 title = {Deep Leakage from Gradients},
 url = {https://proceedings.neurips.cc/paper_files/paper/2019/file/60a6c4002cc7b29142def8871531281a-Paper.pdf},
 volume = {32},
 year = {2019}
}

@misc{salem2018mlleaksmodeldataindependent,
      title={ML-Leaks: Model and Data Independent Membership Inference Attacks and Defenses on Machine Learning Models}, 
      author={Ahmed Salem and Yang Zhang and Mathias Humbert and Pascal Berrang and Mario Fritz and Michael Backes},
      year={2018},
      eprint={1806.01246},
      archivePrefix={arXiv},
      primaryClass={cs.CR},
      url={https://arxiv.org/abs/1806.01246}, 
}

@InProceedings{pmlr-v235-zarifzadeh24a,
  title = 	 {Low-Cost High-Power Membership Inference Attacks},
  author =       {Zarifzadeh, Sajjad and Liu, Philippe and Shokri, Reza},
  booktitle = 	 {Proceedings of the 41st International Conference on Machine Learning},
  pages = 	 {58244--58282},
  year = 	 {2024},
  editor = 	 {Salakhutdinov, Ruslan and Kolter, Zico and Heller, Katherine and Weller, Adrian and Oliver, Nuria and Scarlett, Jonathan and Berkenkamp, Felix},
  volume = 	 {235},
  series = 	 {Proceedings of Machine Learning Research},
  month = 	 {21--27 Jul},
  publisher =    {PMLR},
  url = 	 {https://proceedings.mlr.press/v235/zarifzadeh24a.html}
}

@inproceedings{mattern-etal-2023-membership,
    title = "Membership Inference Attacks against Language Models via Neighbourhood Comparison",
    author = {Mattern, Justus  and
      Mireshghallah, Fatemehsadat  and
      Jin, Zhijing  and
      Sch{\"o}lkopf, Bernhard  and
      Sachan, Mrinmaya  and
      Berg-Kirkpatrick, Taylor},
    editor = "Rogers, Anna  and
      Boyd-Graber, Jordan  and
      Okazaki, Naoaki",
    booktitle = "Findings of the {A}ssociation for {C}omputational {L}inguistics: {ACL} 2023",
    month = jul,
    year = "2023",
    address = "Toronto, Canada",
    publisher = "Association for Computational Linguistics",
    url = "https://aclanthology.org/2023.findings-acl.719/",
    doi = "10.18653/v1/2023.findings-acl.719",
    pages = "11330--11343"
}

@misc{shi2024detectingpretrainingdatalarge,
      title={Detecting Pretraining Data from Large Language Models}, 
      author={Weijia Shi and Anirudh Ajith and Mengzhou Xia and Yangsibo Huang and Daogao Liu and Terra Blevins and Danqi Chen and Luke Zettlemoyer},
      year={2024},
      eprint={2310.16789},
      archivePrefix={arXiv},
      primaryClass={cs.CL},
      url={https://arxiv.org/abs/2310.16789}, 
}

@article{TEIXEIRA2026101812,
title = {Exploring features for membership inference in ASR model auditing},
journal = {Computer Speech \& Language},
volume = {95},
pages = {101812},
year = {2026},
issn = {0885-2308},
doi = {https://doi.org/10.1016/j.csl.2025.101812},
url = {https://www.sciencedirect.com/science/article/pii/S0885230825000373},
author = {Francisco Teixeira and Karla Pizzi and Raphaël Olivier and Alberto Abad and Bhiksha Raj and Isabel Trancoso}
}

@inproceedings{tseng22_interspeech,
  title     = {{Membership Inference Attacks Against Self-supervised Speech Models}},
  author    = {Wei-Cheng Tseng and Wei-Tsung Kao and Hung-yi Lee},
  year      = {2022},
  booktitle = {{Interspeech 2022}},
  pages     = {5040--5044},
  doi       = {10.21437/Interspeech.2022-11245},
  issn      = {2958-1796},
}

@inproceedings{10.1145/3209978.3210025,
author = {Qin, Chuan and Zhu, Hengshu and Xu, Tong and Zhu, Chen and Jiang, Liang and Chen, Enhong and Xiong, Hui},
title = {Enhancing Person-Job Fit for Talent Recruitment: An Ability-aware Neural Network Approach},
year = {2018},
isbn = {9781450356572},
publisher = {Association for Computing Machinery},
address = {New York, NY, USA},
url = {https://doi.org/10.1145/3209978.3210025},
doi = {10.1145/3209978.3210025},
booktitle = {The 41st International ACM SIGIR Conference on Research \& Development in Information Retrieval},
pages = {25–34},
numpages = {10},
location = {Ann Arbor, MI, USA},
series = {SIGIR '18}
}

@inproceedings{10.1145/3626772.3657776,
author = {Zheng, Zhi and Hu, Xiao and Gao, Shanshan and Zhu, Hengshu and Xiong, Hui},
title = {MIRROR: A Multi-View Reciprocal Recommender System for Online Recruitment},
year = {2024},
isbn = {9798400704314},
publisher = {Association for Computing Machinery},
address = {New York, NY, USA},
url = {https://doi.org/10.1145/3626772.3657776},
doi = {10.1145/3626772.3657776},
booktitle = {Proceedings of the 47th International ACM SIGIR Conference on Research and Development in Information Retrieval},
pages = {543–552},
numpages = {10},
location = {Washington DC, USA},
series = {SIGIR '24}
}

@inproceedings{10.1145/3219819.3219823,
author = {Zhou, Guorui and Zhu, Xiaoqiang and Song, Chenru and Fan, Ying and Zhu, Han and Ma, Xiao and Yan, Yanghui and Jin, Junqi and Li, Han and Gai, Kun},
title = {Deep Interest Network for Click-Through Rate Prediction},
year = {2018},
isbn = {9781450355520},
publisher = {Association for Computing Machinery},
address = {New York, NY, USA},
url = {https://doi.org/10.1145/3219819.3219823},
doi = {10.1145/3219819.3219823},
booktitle = {Proceedings of the 24th ACM SIGKDD International Conference on Knowledge Discovery \& Data Mining},
pages = {1059–1068},
numpages = {10},
location = {London, United Kingdom},
series = {KDD '18}
}

@ARTICLE{1167344,
  author={Linden, G. and Smith, B. and York, J.},
  journal={IEEE Internet Computing}, 
  title={Amazon.com recommendations: item-to-item collaborative filtering}, 
  year={2003},
  volume={7},
  number={1},
  pages={76-80},
  doi={10.1109/MIC.2003.1167344}}

@inproceedings{10.1145/2959100.2959190,
author = {Covington, Paul and Adams, Jay and Sargin, Emre},
title = {Deep Neural Networks for YouTube Recommendations},
year = {2016},
isbn = {9781450340359},
publisher = {Association for Computing Machinery},
address = {New York, NY, USA},
url = {https://doi.org/10.1145/2959100.2959190},
doi = {10.1145/2959100.2959190},
booktitle = {Proceedings of the 10th ACM Conference on Recommender Systems},
pages = {191–198},
numpages = {8},
location = {Boston, Massachusetts, USA},
series = {RecSys '16}
}

@article{10.1145/2843948,
author = {Gomez-Uribe, Carlos A. and Hunt, Neil},
title = {The Netflix Recommender System: Algorithms, Business Value, and Innovation},
year = {2016},
issue_date = {January 2016},
publisher = {Association for Computing Machinery},
address = {New York, NY, USA},
volume = {6},
number = {4},
issn = {2158-656X},
url = {https://doi.org/10.1145/2843948},
doi = {10.1145/2843948},
journal = {ACM Trans. Manage. Inf. Syst.},
month = dec,
articleno = {13},
numpages = {19}
}

@inproceedings{10.1145/3298689.3347037,
author = {Wen, Hongyi and Yang, Longqi and Estrin, Deborah},
title = {Leveraging post-click feedback for content recommendations},
year = {2019},
isbn = {9781450362436},
publisher = {Association for Computing Machinery},
address = {New York, NY, USA},
url = {https://doi.org/10.1145/3298689.3347037},
doi = {10.1145/3298689.3347037},
booktitle = {Proceedings of the 13th ACM Conference on Recommender Systems},
pages = {278–286},
numpages = {9},
location = {Copenhagen, Denmark},
series = {RecSys '19}
}

@inproceedings{10.1145/3460231.3474274,
author = {Damak, Khalil and Khenissi, Sami and Nasraoui, Olfa},
title = {Debiased Explainable Pairwise Ranking from Implicit Feedback},
year = {2021},
isbn = {9781450384582},
publisher = {Association for Computing Machinery},
address = {New York, NY, USA},
url = {https://doi.org/10.1145/3460231.3474274},
doi = {10.1145/3460231.3474274},
booktitle = {Proceedings of the 15th ACM Conference on Recommender Systems},
pages = {321–331},
numpages = {11},
location = {Amsterdam, Netherlands},
series = {RecSys '21}
}

@inproceedings{10.1145/3109859.3109906,
author = {Yagci, Murat and Aytekin, Tevfik and Gurgen, Fikret},
title = {On Parallelizing SGD for Pairwise Learning to Rank in Collaborative Filtering Recommender Systems},
year = {2017},
isbn = {9781450346528},
publisher = {Association for Computing Machinery},
address = {New York, NY, USA},
url = {https://doi.org/10.1145/3109859.3109906},
doi = {10.1145/3109859.3109906},
booktitle = {Proceedings of the Eleventh ACM Conference on Recommender Systems},
pages = {37–41},
numpages = {5},
location = {Como, Italy},
series = {RecSys '17}
}

@inproceedings{10.1145/3604915.3610236,
author = {Wilm, Timo and Normann, Philipp and Baumeister, Sophie and Kobow, Paul-Vincent},
title = {Scaling Session-Based Transformer Recommendations using Optimized Negative Sampling and Loss Functions},
year = {2023},
isbn = {9798400702419},
publisher = {Association for Computing Machinery},
address = {New York, NY, USA},
url = {https://doi.org/10.1145/3604915.3610236},
doi = {10.1145/3604915.3610236},
booktitle = {Proceedings of the 17th ACM Conference on Recommender Systems},
pages = {1023–1026},
numpages = {4},
location = {Singapore, Singapore},
series = {RecSys '23}
}

@inproceedings{10.1145/3640457.3688064,
author = {Boratto, Ludovico and Fabbri, Francesco and Fenu, Gianni and Marras, Mirko and Medda, Giacomo},
title = {Fair Augmentation for Graph Collaborative Filtering},
year = {2024},
isbn = {9798400705052},
publisher = {Association for Computing Machinery},
address = {New York, NY, USA},
url = {https://doi.org/10.1145/3640457.3688064},
doi = {10.1145/3640457.3688064},
booktitle = {Proceedings of the 18th ACM Conference on Recommender Systems},
pages = {158–168},
numpages = {11},
location = {Bari, Italy},
series = {RecSys '24}
}

@inproceedings{10.1145/3488560.3498477,
author = {Fazelnia, Ghazal and Simon, Eric and Anderson, Ian and Carterette, Benjamin and Lalmas, Mounia},
title = {Variational User Modeling with Slow and Fast Features},
year = {2022},
isbn = {9781450391320},
publisher = {Association for Computing Machinery},
address = {New York, NY, USA},
url = {https://doi.org/10.1145/3488560.3498477},
doi = {10.1145/3488560.3498477},
booktitle = {Proceedings of the Fifteenth ACM International Conference on Web Search and Data Mining},
pages = {271–279},
numpages = {9},
location = {Virtual Event, AZ, USA},
series = {WSDM '22}
}

@InProceedings{pmlr-v139-kaya21a,
  title = 	 {When Does Data Augmentation Help With Membership Inference Attacks?},
  author =       {Kaya, Yigitcan and Dumitras, Tudor},
  booktitle = 	 {Proceedings of the 38th International Conference on Machine Learning},
  pages = 	 {5345--5355},
  year = 	 {2021},
  editor = 	 {Meila, Marina and Zhang, Tong},
  volume = 	 {139},
  series = 	 {Proceedings of Machine Learning Research},
  month = 	 {18--24 Jul},
  publisher =    {PMLR},
  url = 	 {https://proceedings.mlr.press/v139/kaya21a.html}
}

@inproceedings{zeng-etal-2025-mitigating,
    title = "Mitigating the Privacy Issues in Retrieval-Augmented Generation ({RAG}) via Pure Synthetic Data",
    author = "Zeng, Shenglai  and
      Zhang, Jiankun  and
      He, Pengfei  and
      Ren, Jie  and
      Zheng, Tianqi  and
      Lu, Hanqing  and
      Xu, Han  and
      Liu, Hui  and
      Xing, Yue  and
      Tang, Jiliang",
    editor = "Christodoulopoulos, Christos  and
      Chakraborty, Tanmoy  and
      Rose, Carolyn  and
      Peng, Violet",
    booktitle = "Proceedings of the 2025 Conference on Empirical Methods in Natural Language Processing",
    month = nov,
    year = "2025",
    address = "Suzhou, China",
    publisher = "Association for Computational Linguistics",
    url = "https://aclanthology.org/2025.emnlp-main.1247/",
    doi = "10.18653/v1/2025.emnlp-main.1247",
    pages = "24527--24558",
    ISBN = "979-8-89176-332-6"
}

@misc{kulkarni2026membershipinferenceattacksretrieval,
      title={Membership Inference Attacks for Retrieval Based In-Context Learning for Document Question Answering}, 
      author={Tejas Kulkarni and Antti Koskela and Laith Zumot},
      year={2026},
      eprint={2605.04116},
      archivePrefix={arXiv},
      primaryClass={cs.CR},
      url={https://arxiv.org/abs/2605.04116}, 
}

@inproceedings{3780338.3783063,
author = {Xiang, Zhen and Zheng, Linzhi and Li, Yanjie and Hong, Junyuan and Li, Qinbin and Xie, Han and Zhang, Jiawei and Xiong, Zidi and Xie, Chulin and Yang, Carl and Song, Dawn and Li, Bo},
title = {GuardAgent: safeguard LLM agents via knowledge-enabled reasoning},
year = {2025},
publisher = {JMLR.org},
booktitle = {Proceedings of the 42nd International Conference on Machine Learning},
articleno = {2725},
numpages = {27},
location = {Vancouver, Canada},
series = {ICML'25}
}

@inproceedings{wang-etal-2025-privacy,
    title = "Privacy in Action: Towards Realistic Privacy Mitigation and Evaluation for {LLM}-Powered Agents",
    author = "Wang, Shouju  and
      Yu, Fenglin  and
      Liu, Xirui  and
      Qin, Xiaoting  and
      Zhang, Jue  and
      Lin, Qingwei  and
      Zhang, Dongmei  and
      Rajmohan, Saravan",
    editor = "Christodoulopoulos, Christos  and
      Chakraborty, Tanmoy  and
      Rose, Carolyn  and
      Peng, Violet",
    booktitle = "Findings of the {A}ssociation for {C}omputational {L}inguistics: {EMNLP} 2025",
    month = nov,
    year = "2025",
    address = "Suzhou, China",
    publisher = "Association for Computational Linguistics",
    url = "https://aclanthology.org/2025.findings-emnlp.925/",
    doi = "10.18653/v1/2025.findings-emnlp.925",
    pages = "17055--17074",
    ISBN = "979-8-89176-335-7"
}

@inproceedings{3172077.3172231,
author = {Man, Tong and Shen, Huawei and Jin, Xiaolong and Cheng, Xueqi},
title = {Cross-domain recommendation: an embedding and mapping approach},
year = {2017},
isbn = {9780999241103},
publisher = {AAAI Press},
booktitle = {Proceedings of the 26th International Joint Conference on Artificial Intelligence},
pages = {2464–2470},
numpages = {7},
location = {Melbourne, Australia},
series = {IJCAI'17}
}

@inproceedings{10.1145/1401890.1401969,
author = {Singh, Ajit P. and Gordon, Geoffrey J.},
title = {Relational learning via collective matrix factorization},
year = {2008},
isbn = {9781605581934},
publisher = {Association for Computing Machinery},
address = {New York, NY, USA},
url = {https://doi.org/10.1145/1401890.1401969},
doi = {10.1145/1401890.1401969},
booktitle = {Proceedings of the 14th ACM SIGKDD International Conference on Knowledge Discovery and Data Mining},
pages = {650–658},
numpages = {9},
location = {Las Vegas, Nevada, USA},
series = {KDD '08}
}

@InProceedings{10.1007/978-3-642-40991-2_11,
author="Gao, Sheng
and Luo, Hao
and Chen, Da
and Li, Shantao
and Gallinari, Patrick
and Guo, Jun",
editor="Blockeel, Hendrik
and Kersting, Kristian
and Nijssen, Siegfried
and {\v{Z}}elezn{\'y}, Filip",
title="Cross-Domain Recommendation via Cluster-Level Latent Factor Model",
booktitle="Machine Learning and Knowledge Discovery in Databases",
year="2013",
publisher="Springer Berlin Heidelberg",
address="Berlin, Heidelberg",
pages="161--176",
isbn="978-3-642-40991-2"
}

\end{document}